\documentclass[fleqn,usenatbib]{mnras}

\usepackage{mathptmx}
\usepackage[T1]{fontenc}

\DeclareRobustCommand{\VAN}[3]{#2}
\let\VANthebibliography\thebibliography
\def\thebibliography{\DeclareRobustCommand{\VAN}[3]{##3}\VANthebibliography}

\usepackage{graphicx}	
\usepackage{amsmath}	
\usepackage{graphicx,color,pdfpages,amsmath,float,subfig,amssymb}
\usepackage{makecell}
\usepackage{chemformula}
\usepackage[table, dvipsnames]{xcolor}
\usepackage{booktabs}
\usepackage{hyperref}
\defcitealias{Partridge}{Hu\v{s}ko et al.\ (in prep.)}

\newcommand{\msun}{{\rm{M}_\odot}}
\newcommand{\HI}{HI}

\newcommand{\eagle}{\textsc{eagle}}  
\newcommand{\simba}{\textsc{simba}}  
\newcommand{\gimic}{\textsc{gimic}}  
\newcommand{\colibre}{\textsc{colibre}}
\newcommand{\chimes}{\textsc{chimes}}
\newcommand{\hchimes}{\textsc{hybrid-chimes}}
\newcommand{\sphenix}{\textsc{Sphenix}}
\newcommand{\swift}{\textsc{Swift}}
\newcommand{\monofonic}{\textsc{monofonIC}}
\newcommand{\HBTherons}{\textsc{HBT-HERONS}}
\newcommand{\HBTplus}{\textsc{HBT+}}
\newcommand{\sersic}{S\'{e}rsic}

\title[\textsc{COLIBRE}: galaxy size and angular momentum]{The evolution of the sizes and angular momentum content of galaxies in the COLIBRE simulations}

\author[A. D. Ludlow et al.]{
  Aaron D. Ludlow,$^{1}$\thanks{E-mail: aaron.ludlow@uwa.edu.au}
  Katy L. Proctor,$^{2}$
  Joop Schaye,$^{3}$
  Filip Hu\v{s}ko,$^{3}$ 
  Victor J. Forouhar Moreno,$^{3}$ \newauthor
  Danail Obreschkow,$^{1}$
  Evgenii Chaikin,$^{3,8}$ 
  Matthieu Schaller,$^{3,4}$
  Sylvia Ploeckinger,$^{5}$
  Alejandro Ben\'{i}tez-Llambay,$^{7}$ \newauthor
  Kyle A. Oman,$^{8,9}$
  Robert J. McGibbon,$^{3}$
  James W. Trayford,$^{6}$ 
  Carlos S. Frenk,$^{8}$
  Alexander J. Richings$^{10,11}$ \\
  $^{1}$International Centre for Radio Astronomy Research, University of Western Australia, 35 Stirling Highway, Crawley, Western Australia, 6009, Australia \\
  $^{2}$The Oskar Klein Centre, Department of Physics, Stockholm University, AlbaNova University Center, 106 91 Stockholm, Sweden \\
  $^{3}$Leiden Observatory, Leiden University, PO Box 9513, 2300 RA Leiden, the Netherlands \\
  $^{4}$Lorentz Institute for Theoretical Physics, Leiden University, PO Box 9506, 2300 RA Leiden, the Netherlands \\
  $^{5}$Department of Astrophysics, University of Vienna, T\"{u}rkenschanzstrasse 17, 1180 Vienna, Austria \\
  $^{6}$Institute of Cosmology and Gravitation, University of Portsmouth, Dennis Sciama Building, Burnaby Road, Portsmouth PO1 3FX, UK \\
  $^{7}$Dipartimento di Fisica G. Occhialini, Universit\`{a} degli Studi di Milano Bicocca, Piazza della Scienza, 3 I-20126 Milano MI, Italy \\
  $^{8}$Institute for Computational Cosmology, Department of Physics, University of Durham, South Road, Durham, DH1 3LE, UK \\
  $^{9}$Centre for Extragalactic Astronomy, Department of Physics, University of Durham, South Road, Durham, DH1 3LE, UK \\
  $^{10}$Centre for Data Science, Artificial Intelligence and Modelling, University of Hull, Cottingham Road, Hull, HU6 7RX, UK \\
  $^{11}$E. A. Milne Centre for Astrophysics, University of Hull, Cottingham Road, Hull, HU6 7RX, UK \\
}

\date{Accepted XXX. Received YYY; in original form ZZZ}
\pubyear{2026}

\begin{document}
\label{firstpage}
\pagerange{\pageref{firstpage}--\pageref{lastpage}}
\maketitle

\begin{abstract}
  We analyse the sizes and specific angular momentum content of galaxies in the \textsc{Colibre} cosmological hydrodynamical
  simulations spanning two orders of magnitude in mass resolution. We compare the predicted size--mass and angular momentum--mass
  relations to a broad range of observational
  measurements spanning redshifts $z=0$ to $4$. At $z=0$, \textsc{Colibre} reproduces observed size--mass relations
  over the sampled mass range $10^8 \lesssim M_\star/{\rm M_\odot}\lesssim 10^{11.5}$, and for multiple size definitions, including two- and
  three-dimensional stellar half-mass radii, half-light radii across several wavelengths, as well as alternative measures
  such as baryonic half-mass radii and characteristic radii defined by stellar surface density thresholds. The simulations
  also recover the observed segregation of galaxies in the size--mass plane by morphological type and star formation rate,
  and reproduce the distinct, approximately parallel sequences followed by star-forming discs and quenched spheroids in the
  stellar specific angular momentum--mass plane. The angular momentum content of star-forming \colibre~galaxies matches that of observed
  systems out to $z\approx 1.5$. At higher
  redshifts, massive galaxies ($ 10^{9.5}\lesssim M_\star/{\rm M_\odot}\lesssim 10^{11}$) in the simulations are somewhat
  smaller than observed, and the separation between star-forming and passive populations in the size--mass plane is
  reduced relative to observations, while at lower masses the agreement remains good. This apparent discrepancy may
  reflect the effects of dust attenuation, which is neglected in our analysis and may preferentially obscure
  the central regions of observed systems. Overall, our findings highlight the close connection between galaxy size, angular
  momentum, and morphology over cosmic time.
\end{abstract}

\begin{keywords}
  methods: numerical -- galaxies: formation, evolution
\end{keywords}

\section{Introduction}
\label{SecIntro}

The stellar mass ($M_\star$) and size ($R_\star$) of galaxies are among their most fundamental properties. 
These quantities are strongly correlated and vary systematically with morphology, star-formation activity, and colour, 
among other properties \citep[e.g.][]{Shen2003, vanderWel2014, vandokkum2015, Lange2015, Whitaker2017, Cook2025}. 
As a result, considerable effort has been devoted to characterising the galaxy size--mass relation (SMR) and
its dependence on secondary galaxy parameters in order to constrain models of galaxy formation and evolution.

There is now broad consensus from observational studies on a few key features of the SMR:  
(1) At fixed mass, late-type discs are typically larger, bluer, and more actively star-forming than early-type spheroids,
which are generally smaller, redder, and quenched \citep[e.g.][]{Shen2003,Trujillo2007,Whitaker2017};
(2) At fixed redshift, galaxy size increases on average with stellar mass, while at fixed stellar mass it decreases with
redshift \citep[e.g.][]{vanderWel2014,Ward2024}; and (3) size evolution is stronger for massive and quenched galaxy
populations than for low-mass or star-forming ones, although individual galaxies may transition between these
populations \citep{Trujillo2007,vanderWel2014,Mowla2019}.

The SMR also depends on the underlying distribution of galaxy morphologies. At $z\approx 0$,
spheroid-dominated early-type galaxies follow a steep SMR with a logarithmic slope between $\approx 0.4$ and 0.6, while
late-type discs exhibit a shallower slope of $\approx 0.3$ \citep{Shen2003, Lange2015}. Since early-types dominate the
population above $M_\star\approx 10^{10.5} {\rm M}_{\odot }$---whereas lower-mass systems comprise a more diverse range of
morphologies---the global SMR is observed to steepen above this mass threshold.

Although more difficult to directly observe, the stellar specific angular momentum ($j_\star$) provides a complementary
probe of galaxy structure. 
Early work by \citet{Fall1983} demonstrated that nearby spiral and elliptical galaxies occupy approximately parallel sequences 
in the $\log j_\star - \log M_\star$ plane (the angular momentum--mass relation, or AMR for short)
with slopes of $\approx 0.7$, but with spheroids offset to lower angular momentum
at fixed mass by $\approx 0.7$ dex. \citet{FR2013} confirmed these relations using an updated sample of galaxies that span
a broader range of morphological types. They reported similar slopes but larger offsets ($\approx 0.9$ dex) between pure discs and
pure spheroids, and a normalisation that depends smoothly on the spheroid-to-total ratio ($S/T$) between these extremes. 
\citet{Huang2017} confirmed that discs are larger than spheroids of the same stellar mass, in a manner that is consistent with the
morphology-dependent normalization of the $j_\star - M_\star$ relation inferred by these earlier studies.

In the standard picture of disc formation, dark matter haloes acquire angular momentum through tidal torques from
surrounding large-scale structure \citep{Hoyle1949,Peebles1969,White1984}. Because baryons and dark matter are initially well mixed, they are expected
to gain similar specific angular momenta at early times. However, simulations
indicate that present-day galaxies retain only a fraction of their initial angular momentum \citep{Frenk1985},
which correlats with the fraction of halo mass retained by galaxies \citep[e.g.][]{Sales2009,Sales2010}.
In hydrodynamical simulations such as TNG100, late-type galaxies retain $\sim$50--60 per cent of their halo's specific
angular momentum, while early types retain only $\sim$10--20 per cent \citep{Genel2015, RG2022}, broadly consistent with
observational estimates of angular momentum retention fractions \citep{FR2013, Posti2018, DiTeodoro2021, FR2018}.
Reproducing both the magnitude of angular momentum retention and its dependence on morphology is therefore essential for
matching observed galaxy sizes.                                                                                                      

Early hydrodynamical simulations of galaxy formation suffered catastrophic angular momentum losses and produced galaxies
that were too small and had too little angular momentum when compared to observed systems \citep[e.g.][]{Navarro1995b, NS1997}.
This shortcoming was later attributed to limited resolution and ineffective or absent stellar feedback
\citep[e.g.][]{Thacker2001, MallerDekel2002}, which allowed excessive radiative cooling and efficient star formation in galaxy progenitors.
Mergers between these dense stellar clumps transferred orbital angular momentum to the halo, reducing $j_\star$ and leading to
overly compact galaxies.

The problem was alleviated in subsequent higher-resolution simulations that incorporated efficient stellar and active galactic nucleus (AGN) feedback.
\citep[e.g.][]{Governato2004, Governato2007, Okamoto2005, Scannapieco2008a}. By reheating or expelling gas, such feedback delays
star formation and suppresses early angular momentum loss. All modern simulations now include these processes, and some---though
not all---successfully reproduce the observed sizes and angular momentum distributions of galaxies, as well as their dependence on
morphology, colour, and star formation rate \citep[e.g.][]{Schaye2015, Teklu2015, Snyder2015, Furlong2017, RG2022}.

In the \gimic~simulations \citep{Crain2009}, galaxies with
$10^9 \lesssim M_\star/{\rm M_\odot} \lesssim 10^{10.5}$ have realistic sizes and rotation velocities
\citep{McCarthy2012}, whereas more massive systems are overly compact and excessively star forming, likely
due the absence of feedback from supermassive black holes, which is known to affect the properties of massive systems.
Galaxies in the original Illustris simulation \citep{Vogelsberger2014} also display a SMR that is systematic offset from the observed relation,
with galaxies of $M_\star \lesssim 10^{11}\,{\rm M_\odot}$ being larger than observed by roughly a factor of 2 \citep{Snyder2015}.
\citet{Dubois2016} showed that the sizes of star-forming galaxies in the Horizon-AGN simulation \citep{Dubois2014} are broadly consistent with
observations, but quiescent systems below $M_\star \approx 10^{11.3}\,{\rm M_\odot}$ are larger than star-forming ones,
inconsistent with what is observed. A similar trend in the size of star-forming and passive galaxies has been reported for
\simba~\citep{Dave2019}.

Yet a number of simulations show good agreement with observations. The \eagle~simulation \citep{Furlong2017}, in particular, reproduces
the $z=0$ SMR of \citet{vanderWel2014} to within $\approx 0.1$ dex for both star-forming
and passive galaxies, and matches the evolution of galaxy sizes from $z\approx 2$ to $z\approx 0$ to within $\approx 0.2$
dex. Similarly, the Magneticum-Pathfinder simulations produce a galaxy population with size distributions consistent with observations,
particularly for early-type galaxies \citep{Schulze2018,Dolag2025}.

Simulations that broadly match the SMR and its dependence on galaxy type also tend to reproduce the observed AMR. This has been
demonstrated for \eagle~\citep{Lagos2017} and Magneticum-Pathfinder \citep{Teklu2015}, where discs and spheroids occupy distinct but
approximately parallel sequences in the $\log j_\star$–$\log M_\star$ plane, as observed, with similar behaviour also found in the Illustris
simulation \citep{Genel2015}.

Despite this progress, quantitative comparisons between simulations and observations remain sensitive to
measurement methodology. In simulations, galaxy sizes are typically measured directly from particle data
within fixed three-dimensional apertures that also define the stellar mass, so the inferred masses and half-mass radii
depend on the adopted aperture, at least for massive systems \citep[e.g.][]{Furlong2017}. In contrast, observational sizes are often 
light-weighted and therefore wavelength dependent, reflecting dust attenuation and spatial variations in the stellar mass-to-light ratio.
Galaxies generally appear larger at shorter wavelengths \citep[e.g.][]{Vulcani2014,Suess2019}, because young
stars preferentially reside at large radii and centrally-concentrated dust attenuates blue light.
To reduce biases when comparing simulations with observations, it is increasingly common to construct mock images of
simulated galaxies by modelling the spectral energy distributions of stellar particles, sometimes including additional effects
such as dust attenuation, background emission, and instrumental noise. Sizes measured from such light-weighted images are
systematically larger than their mass-weighted counterparts \citep[e.g.][]{vdSande2019,deGraaff2022}.

While sizes are traditionally measured using half-light or half-mass radii, alternative definitions have
been proposed that aim to better capture the physical extent of galaxies. For example, \citet{Trujillo2020} introduced the radius
$R_1$, defined by a stellar surface density threshold of $1\,{\rm M_\odot\,pc^{-2}}$. This roughly matches the typical gas density
threshold for star formation \citep[e.g.][]{Schaye2004,Rajohnson2022} and serves as a useful proxy for the “edge” of the in-situ
stellar component \citep[e.g.][]{Chamba2022}.
Meanwhile, \citet{Zichen2025} used the complete census of galactic baryons to measure the ``baryonic size--mass relation''
of galaxies in the Spitzer Photometry and Accurate Rotation Curves sample \citep[SPARC;][]{Lelli2016}.

In this paper, we study the sizes and angular momentum content of galaxies in the
\textsc{Colibre}\footnote{\href{https://colibre.strw.leidenuniv.nl/}{https://colibre.strw.leidenuniv.nl/}}
cosmological hydrodynamical simulations \citep{Schaye2026, Chaikin2026a}, which reproduce a broad range of observed galaxy properties
across cosmic time. These include the stellar mass function, star-formation rates, passive fractions, and the masses of atomic and
molecular gas and dust \citep{Schaye2026,Chaikin2026b}, as well as the resolved Kennicutt--Schmidt relation \citep{Lagos2026}. Focusing
on galaxies at $z \lesssim 4$, we compare the simulated size--mass and $j_\star$--$M_\star$ relations to observational measurements, and
examine their dependence on morphology, star-formation activity, and size definition. Our primary goal is to assess how well
\textsc{Colibre}~reproduces these observed relations across galaxy populations, redshifts, and size definitions. In this respect,
the paper is intentionally more descriptive than theoretical, with a focus on identifying where the model succeeds or fails to reproduce
observations, thereby helping to clarify the physical regimes and galaxy populations for which the simulation provides a robust
description. We consider both standard measures (half-mass and half-light radii) and alternative metrics such as $R_1$ and the
half-baryonic mass radius, $R_{\rm bar,50}$. Results at $z > 4$, and those based on mock observational imaging, including dust,
will be presented elsewhere.

This paper is organised as follows. 
Section~\ref{SSecColibre} describes the \textsc{Colibre} simulations, while Section~\ref{SSecHBT} outlines our halo and galaxy 
identification methods. Our analysis techniques are presented in Section~\ref{SSecAnalysis}. 
Section~\ref{ref:results} presents our main results, including 1) the SMR at $z=0$ (Section~\ref{sec:SM-lowz}) and its evolution with
redshift (Section~\ref{sec:SM-highz}); and 2) the AMR relation at $z=0$ (Section~\ref{sec:low-z-ang}) and its evolution with redshift
(Section~\ref{sec:JM-highz}). We summarise our findings in Section~\ref{Summary}. We validate some of our choices and assumptions
in the appendices, including tests of the minimum stellar particle number (Appendix~\ref{app:convergence}), differences between
central and satellite galaxies (Appendix~\ref{app:centsat}), and the dependence on aperture choice (Appendix~\ref{app:aperture})
and observational scatter in stellar masses (Appendix~\ref{app:scatter}). In Appendix~\ref{app:nsersic}, we compare the \sersic~indices measured
for simulated and observed galaxies. 

\section{Simulations and analysis}
\label{SecS}

\subsection{The \colibre~Simulations}
\label{SSecColibre}

\subsubsection{Cosmology and initial conditions}
\label{SSSecCosmoIC}

Our analysis is based on \colibre, a new suite of cosmological hydrodynamical simulations of galaxy formation \citep{Schaye2026}.
All simulations were run with the \swift~code \citep{Schaller2024}, using the \sphenix~smoothed particle hydrodynamics (SPH) scheme
to model hydrodynamic interactions between gas elements \citep{Borrow2022}.

Initial conditions (ICs) were created at $z=63$ using \monofonic~\citep{Hahn2021,Michaux2021} employing
second order Lagrangian perturbation theory. Cosmological parameters are from the Dark Energy Survey
Year 3 \citep[DES Y3;][]{Abbott2022} ‘3×2pt + All Ext.’ $\Lambda$CDM model.

All of our initial conditions contain four times as many dark matter particles as baryonic ones, whereas nearly all previous cosmological
simulations adopted equal numbers for both species \citep[see][for exceptions]{Tremmel2017,Ludlow2023}. This choice ensures that all
particle species have comparable masses, including stellar particles, which inherit the masses of their gaseous progenitors. As a result,
the effective resolution of stellar structures is increased by suppressing spurious, irreversible energy transfer from dark matter to
stellar particles, which can artificially heat galaxies, inflate their sizes, and disrupt ordered motions\footnote{As discussed by
\citet{Ludlow2021}, the primary benefit of this approach is not that dark matter and stellar particles have similar masses, but that
it increases the number of dark matter particles in haloes where galaxies form. This, in turn, prolongs the local relaxation time and
reduces numerical heating driven by two-body interactions. Disparate particle masses leads to a secondary effect, commonly referred to as
mass segregation.} \citep[e.g.][]{Ludlow2019b,Ludlow2021,Ludlow2023,Wilkinson2023}.

\colibre~features three resolution levels, with particle masses of $\sim 10^5$, $10^6$, and $10^7\, {\rm M_\odot}$, and five box sizes
ranging from $25$ to $400$ comoving ${\rm Mpc}$ (${\rm cMpc}$) in factors of 2. We will distinguish runs using an identifier that encodes
the box size (L) and approximate base-ten logarithm of the particle mass (m); for example, L200m6 refers to the $(200\, {\rm cMpc})^3$
volume simulated with baryonic and dark matter particles of mass $m\sim 10^6\,{\rm M_\odot}$. Gravitational forces between particles
of all types are softened using a \citet{Wendland1995} C2 kernel below a time- and resolution-dependent spatial scale of
$3\times\epsilon$, where $\epsilon$ is the 'Plummer-equivalent' softening length. At $z=0$, these are $\epsilon_{\rm prop}=0.35$,
$0.7$, and $1.4\,{\rm kpc}$ for the m5, m6, and m7 resolutions, respectively. 

In this paper, we mainly present results from the highest resolution runs available for each volume (L025m5 or L050m5, L200m6, and L400m7),
and reserve runs of lower resolution or smaller volumes to assess convergence (see Appendix~\ref{app:convergence}). Table~\ref{TabSimParam}
provides details of the runs, including the box size, the numbers and initial masses of dark matter and baryonic particles, as well as the
maximum co-moving and proper gravitational softening lengths.

\begin{table*}
  \centering 
  \caption{Relevant numerical parameters of the \colibre~simulations. Columns provide a simulation label, the model for active galactic nucleus feedback,
    the comoving box side-length ($L$), the total number of dark matter and baryonic particles ($N_{\rm DM}$ and $N_{\rm bar}$, respectively),
    the initial dark matter and baryonic particle masses ($m_{\rm DM}$ and $m_{\rm bar}$, respectively), and the maximum comoving ($\epsilon_{\rm com}$)
    and physical ($\epsilon_{\rm prop}$) softening lengths.}
  \label{TabSimParam}
  \begin{tabular}{r l r r r r r r r r} 
    \toprule
    & Label & Model & L    & $N_{\rm DM}$ & $N_{\rm bar}$ & $m_{\rm DM}$ & $m_{\rm bar}$ & $\epsilon_{\rm com}$ & $\epsilon_{\rm prop}$ \\
    &       &       & [cMpc] &      &               & [$M_{\odot}$] & [$M_{\odot}$] & [kpc] & [kpc] \\
    \midrule 
    & L400m7  & thermal-AGN & 400 & $4\times 3008^3$ & $3008^3$ & $1.94\times 10^7$  & $1.47\times 10^7$ & 3.6 & 1.4  \\
    & L200m7  & thermal-AGN & 200 & $4\times 1504^3$ & $1504^3$ & $1.94\times 10^7$  & $1.47\times 10^7$ & 3.6 & 1.4  \\
    & L200m6  & thermal-AGN & 200 & $4\times 3008^3$ & $3008^3$ & $2.42\times 10^6$  & $1.84\times 10^6$ & 1.8 & 0.7  \\
    & L100m6  & thermal-AGN & 100 & $4\times 1504^3$ & $1504^3$ & $2.42\times 10^6$  & $1.84\times 10^6$ & 1.8 & 0.7  \\
    & L050m5  & thermal-AGN & 50  & $4\times 1504^3$ & $1504^3$ & $3.03\times 10^5$  & $2.30\times 10^5$ & 0.9 & 0.35 \\
    & L025m7  & thermal-AGN & 25  & $4\times 188^3$  & $188^3$  & $1.94\times 10^7$  & $1.47\times 10^7$ & 3.6 & 1.4  \\
    & L025m6  & thermal-AGN & 25  & $4\times 376^3$  & $376^3$  & $2.42\times 10^6$  & $1.84\times 10^6$ & 1.8 & 0.7  \\
    & L025m5  & thermal-AGN & 25  & $4\times 752^3$  & $752^3$  & $3.03\times 10^5$  & $2.30\times 10^5$ & 0.9 & 0.35 \\
    & L200m7h & hybrid-AGN  & 200 & $4\times 1504^3$ & $1504^3$ & $1.94\times 10^7$  & $1.47\times 10^7$ & 3.6 & 1.4  \\
    & L100m6h & hybrid-AGN  & 100 & $4\times 1504^3$ & $1504^3$ & $2.42\times 10^6$  & $1.84\times 10^6$ & 1.8 & 0.7  \\
    \bottomrule
  \end{tabular}
\end{table*}

\subsubsection{Sub-resolution models and calibration}
\label{SSSecsubgrid}

\colibre~includes sub-resolution models for radiative cooling and heating \citep{Ploeckinger2025}, star formation
\citep{Nobels2024}, black hole growth \citep{Nobels2022,Bahe2022}, dust formation and its role in the regulation of molecular gas
\citep{Trayford2026}, chemical enrichment, and turbulent mixing \citep{Correa2026}. Stellar
feedback is implemented through stellar winds, radiation pressure, HII region heating \citep{BenitezLlambay2025}, and supernovae
\citep[including a stochastic thermal component as well as a kinetic component to drive turbulence;][]{DallaVecchia2012,Chaikin2023}.
\colibre~includes two models for black hole feedback: a purely thermal AGN mode \citep[][thermal-AGN hereafter]{BoothSchaye2009}
and a hybrid thermal/kinetic jet model (hybrid-AGN) that explicitly tracks the spin of the central supermassive black hole
(\citealt{Husko2026}; runs that adopt this feedback scheme will be labelled with an ``h'' suffix; e.g. L200m6h).
The results presented in Sections~\ref{sec:SM-lowz} to~\ref{sec:JM-highz} were obtained from the
thermal-AGN model; we compare them to results from the hybrid-AGN scheme in Section~\ref{sec:hybrid}.

Complementing these feedback and enrichment processes, \colibre’s treatment of the interstellar medium avoids imposing an effective
equation of state, enabling gas to cool radiatively to temperatures $\approx 10\,{\rm K}$.
Gas particles are eligible to form stars when they satisfy a local gravitational instability criterion that depends on gas density,
particle mass, and the local thermal and turbulent velocity dispersions, rather than being explicitly linked to gas temperature.
Consequently, star formation occurs predominantly---but not exclusively---in cold ($T\lesssim 10^4\,{\rm K}$), dense gas with low
velocity dispersion, conditions that are important for accurately modelling galaxy sizes and the vertical structure of galactic
discs. Because the instability criterion is evaluated at the resolution limit of the simulation, the typical densities of
  star-forming gas particles increase with increasing resolution, while their temperatures and turbulent velocities decrease.
To achieve this, \colibre~employs the chemistry solver \chimes~\citep{Richings2014a,Richings2014b} to track the
non-equilibrium abundances of primordial elements, atomic and molecular hydrogen, helium (in all neutral and ionised phases), and dust within each fluid
element, along with their associated heating and cooling rates. Element-by-element metal-line cooling rates are
computed using \hchimes~\citep{Ploeckinger2025} for 9 elements---C, N, O, Ne, Mg, Si, S, Ca, and Fe---assuming chemical equilibrium, but are corrected
to account for the non-equilibrium free electron densities from the hydrogen and helium reaction network.

Subgrid models for stellar and AGN feedback were calibrated at m7 resolution using Gaussian process emulators that were trained
on $\sim 200$ simulations of the same $(50 \, {\rm cMpc})^3$ volume \citep[see][for details]{Chaikin2026a}. Parameters for
m7 runs were obtained by fitting the emulator to the GAMA DR4 $z=0$ galaxy stellar mass function from \citet[][GSMF]{Driver2022}
and to the (projected) SMR of \citet{Hardwick2022}. The fit was performed over the galaxy stellar mass range
$10^9 < M_\star/{\rm M_\odot} < 10^{11.3}$, applying equal weights to both datasets. At the m5 and m6 resolutions, subgrid parameters
were adjusted manually to achieve convergence. The coupling efficiency of AGN feedback was independently calibrated so that
the black hole mass--stellar mass relation agreed with the observed relation for massive galaxies for which robust black hole mass
estimates are available. This calibration strategy yields excellent weak convergence\footnote{As discussed in \citet{Schaye2015}, ‘weak convergence’ refers to
  agreement across resolution levels when subgrid models are recalibrated independently at each resolution. In this approach, subgrid parameters are
  allowed to compensate for changes in galaxy properties arising from different numerical parameters, such as mass and force resolution. Throughout this paper,
  remarks about ``good convergence'' across resolution levels explicitly refer to weak convergence rather than implying insensitivity to unresolved physics.}
in key galaxy properties across all resolution levels.

More details about the calibration and convergence between runs of different resolution can be found in
\citet{Schaye2026}, \citet{Chaikin2026a}, and in Appendix~\ref{app:convergence}.

\subsection{Halo and galaxy identification}
\label{SSecHBT}

Dark matter particles are assigned to top-level haloes using a Friends-of-Friends (FoF)
algorithm (we employ a linking length equal to 0.2 times the mean dark matter inter-particle separation; \citealt{Davis1985});
gas, stellar, and black hole particles are assigned to the same FoF halo as the nearest dark matter particle within
the linking length, provided it belongs to one.

Galaxies and their dark matter haloes were identified using \HBTherons~\citep{HBTHerons}, an updated version of
\HBTplus~\citep{Han2018} suitable for hydrodynamical simulations. An iterative unbinding procedure that accounts for the binding
energies of all particle species was used to decompose FoF haloes into self-bound substructures.
This process leverages the hierarchical formation of structure in $\Lambda$CDM and assumes each subhalo originated as
the central subhalo of an earlier FoF group. Snapshots were thus processed sequentially from early to late times, using 
each subhalo’s particles to identify its descendant in subsequent snapshots. This particle-tracking approach improves subhalo
identification, especially for deeply nested satellites and those embedded in dense regions of their hosts
\citep{HBTHerons,ChandroGomez2025}.

When multiple central subhaloes merge onto a single descendant FoF group between two adjacent snapshots, the
central is assigned to the most massive progenitor, provided no others are within 20 per cent of its mass. Otherwise, the subhalo with the
lowest orbital kinetic energy in the rest frame of the host FoF halo is selected. 

We associate central galaxies with central subhaloes and satellite galaxies with satellite subhaloes, and exclude
the latter from our analysis unless stated otherwise (see Appendix~\ref{app:centsat} for an assessment of how this
choice affects the SMR and AMR).

\subsection{Measurements of galaxy properties}
\label{SSecAnalysis}

The halo and galaxy catalogues were post-processed with the Spherical Overdensity Aperture Processor
\citep[SOAP;][]{McGibbon2025}, which calculates a wide range of subhalo and galaxy properties.
We obtain the stellar mass of each galaxy, $M_\star$, by summing the individual masses of bound stellar
particles that lie within 50 kpc of its centre. Stellar masses defined this way reproduce those inferred from
mock observations of \eagle~galaxies \citep{deGraaff2022}, and unless otherwise stated, this definition is adopted
in all figures that follow.

To obtain specific star formation rates (sSFRs) we sum the instantaneous SFRs of bound gas particles within a 50 kpc aperture
and divide by $M_\star$. Galaxy morphologies are quantified using the kinematic disc-to-total mass fraction, $D/T$. To estimate
$D/T$, we double the mass of bound stellar particles within 50 kpc that counter-rotate with respect to their net rotation 
to approximate the
bulge component, subtract this mass from the total stellar mass to obtain the disc mass, and divide it by $M_\star$. 
The resulting $D/T$ values correlate strongly with other kinematic morphology indicators, such as the co-rotating stellar mass
fraction, $\kappa_{\rm co}$ \citep[e.g.][]{Sales2010,Correa2017}, and the ratio of azimuthal velocity to velocity dispersion \citep{Victor2026}.
  
The 50 kpc apertures\footnote{See
Appendix~\ref{app:aperture} for a comparison of the SMRs and AMRs obtained for different choices of the spherical aperture.}
used to calculate $M_\star$, sSFR, and $D/T$ are centred on the most-bound particle (of any type). All galaxy
properties presented hereafter are defined in a proper coordinate system.

\begin{figure}
  \subfloat{\includegraphics[width=0.48\textwidth]{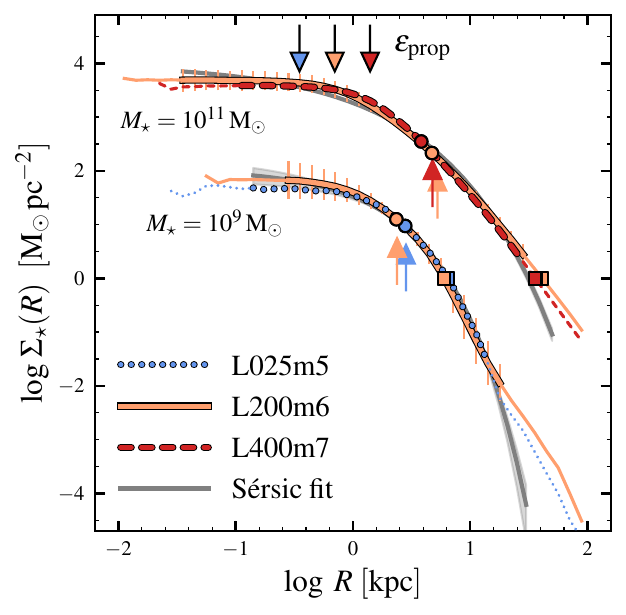}}
  \caption{Median stellar surface density profiles, $\Sigma_\star(R)$, for $z=0$ central galaxies of mass $M_\star = 10^9\, {\rm M_\odot}$ and
    $10^{11}\, {\rm M_\odot}$. Different coloured curves correspond to different simulations. Thick segments mark the radial range
    used for profile fitting and error bars indicate the $16^{\rm th}$ to $84^{\rm th}$ percentile scatter in $\Sigma_\star(R)$ for
    individual galaxies in L200m6. Grey lines show the mean of the best-fitting S\'{e}rsic profiles obtained separately for each simulation and
    mass bin; shaded grey regions span the range corresponding to individual fits to each of the median $\Sigma_\star(R)$ profiles.
    Coloured circles show median $R_{\star,50}$ values measured directly from the particle data, and upward arrows indicate the values
    inferred from the best-fitting S\'{e}rsic profiles. Coloured squares correspond to the radius $R_1$ at which
    $\Sigma_\star=1\,{\rm M_\odot\, pc^{-2}}$. Downward arrows mark the $z=0$ gravitational softening lengths, $\epsilon_{\rm prop}$,
    for each simulation, which exceed the minimum radii where profiles from runs with different resolutions converge.}
  \label{fig:Sigma}
\end{figure}

\subsubsection{Galaxy sizes}
\label{SSSecSizes}

In Sections~\ref{sec:SM-lowz} and \ref{sec:SM-highz}, we compare the sizes of \textsc{Colibre} galaxies with several observational
datasets. Our approach mimics standard observational analysis methods, although we defer
full forward-modelling of the simulated galaxies to future work. Unless stated otherwise, we include only stellar particles bound to
each galaxy when estimating its size, and we have verified that our results are not strongly affected by this choice.

In what follows, we consider several measures of galaxy size motivated by these observational datasets.
One is the three-dimensional stellar
half-mass radius, $r_{\star,50}$, defined as the radius of a sphere centred on the galaxy that encloses half of $M_\star$.
This simple measure provides a clear indication of a galaxy’s physical extent and how it evolves over cosmic time.

We also compute several two-dimensional size measures directly from the particle data: the stellar half-mass radius, $R_{\star,50}$, and the
$u$-, $r$-, and $z$-band\footnote{\colibre~stores the rest-frame, dust-free AB magnitudes of star particles in the GAMA bands,
computed following \citet{Trayford2015}. These were calculated from \citet{BruzualCharlot2003} (GALAXEV) stellar population
synthesis models convolved with the relevant filter transmission curves, and depend on each particle’s age and metallicity.}
half-light radii, denoted $R_{u,50}$, $R_{r,50}$, and $R_{z,50}$, respectively. For the $u$- and $z$-bands, we also compute the
radii enclosing 90 per cent of the light, $R_{u,90}$ and $R_{z,90}$, respectively. 

We also estimate $R^{\rm Ser}_{\star,50}$ and $R^{\rm Ser}_{r,50}$ by fitting single-component \citet{Sersic1963} profiles to the
stellar‐mass and $r$-band surface-density profiles ($\Sigma_\star(R)$ and $\Sigma_r(R)$, respectively). Both
$\Sigma_\star(R)$ and $\Sigma_r(R)$ are initially computed over the radial range $-2\leq \log (R/{\rm kpc}) \leq 3$ in logarithmic
radial bins of width $\Delta\log R=0.1$. We then fit the profiles over a restricted range
$r_{\rm min} \le r \le r_{\rm max}$, where $r_{\rm min}$ is the innermost bin that encloses 10 stellar particles.
The outermost bin, $r_{\rm max}$, is the smaller of either $50\,{\rm kpc}$, or the farthest bin that contains at
least 10 stellar particles.\footnote{We have verified that
the values of $R^{\rm Ser}_{\star,50}$ and $R^{\rm Ser}_{r,50}$ obtained this way are in good agreement with those obtained directly from the
particle data. Note that we use the superscript ``Ser'' to denote half-mass sizes estimated from S\'ersic fits.} These measurements use three orthogonal projections of each galaxy (along the simulation’s
native $x$-, $y$-, and $z$-axes); the two-dimensional sizes presented below are averages over these orientations.

Fig.~\ref{fig:Sigma} shows median $\Sigma_\star(R)$ profiles for galaxies within $0.3$ dex of $M_\star = 10^9\, {\rm M_\odot}$ and
$10^{11}\, {\rm M}_\odot$. Coloured lines denote different simulations; error bars indicate the $16^{\rm th}$ to $84^{\rm th}$
percentile scatter among individual galaxies in each mass bin, and are shown only for the L200m6 run for clarity.
Grey lines show the mean best-fitting S\'{e}rsic profiles
in each mass bin, obtained by averaging fits to the median profiles at fixed mass; grey shading indicates the range of
the individual fits. Thick line segments mark the radial range used for fitting. Note that the 2D half-mass radii inferred from the
S\'{e}rsic fits (upward arrows) agree well with those measured directly from the particle data (coloured circles). Note too that the
profiles obtained for the different resolution runs are in good agreement down to radii that are much smaller than the gravitational softening
lengths ($\epsilon_{\rm prop}$; marked by downward pointing arrows), suggesting that softening is not necessarily a limiting factor
in resolving galaxy sizes \citep[see also][]{Ludlow2023,Ludlow2020}. This justifies the choice of $r_{\rm min}$ used for profile fitting, which is
typically smaller than the softening length.

\begin{figure*}
  \subfloat{\includegraphics[width=0.5\textwidth]{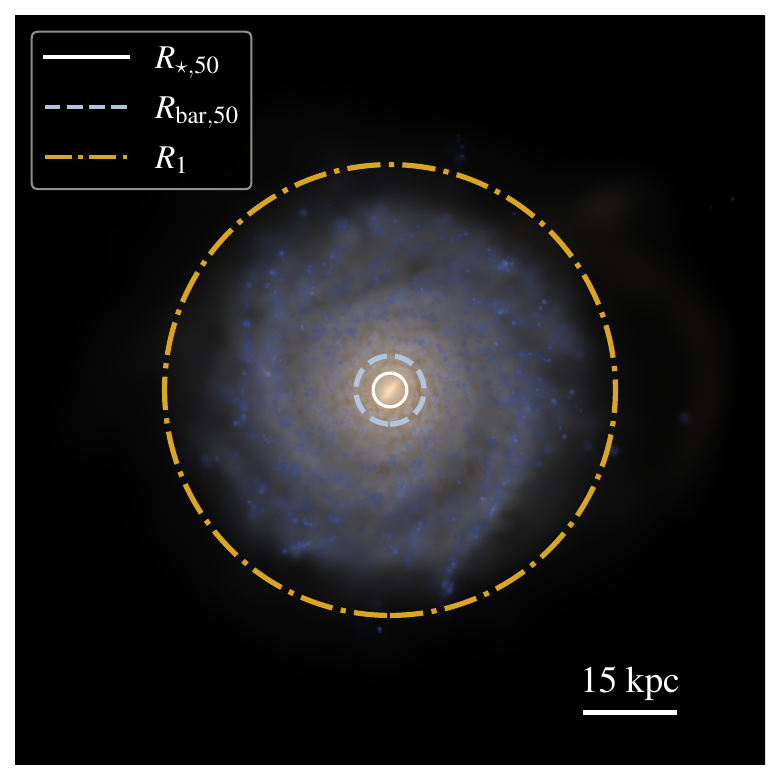}}
  \subfloat{\includegraphics[width=0.5\textwidth]{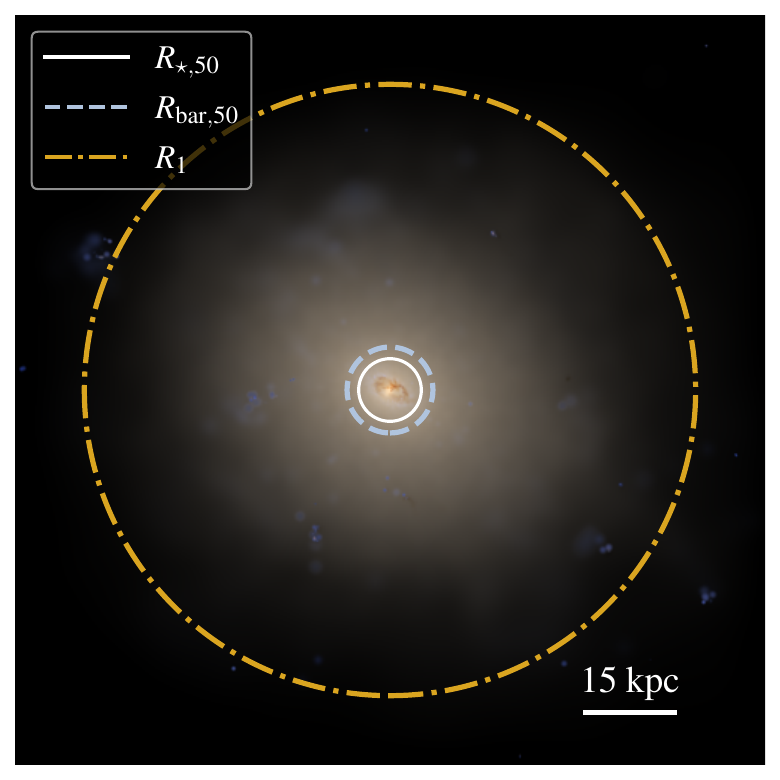}}
  \caption{Virtual \textit{Hubble Space Telescope} observations of a $z=0$ star-forming disc galaxy (left panel;
    $M_\star=8.8\times 10^{10}\,{\rm M_\odot}$, SFR=$2.5\,{\rm M_\odot\, yr^{-1}}$) and a passive elliptical galaxy
    (right panel; $M_\star=1.5\times 10^{11}\,{\rm M_\odot}$, SFR=$0.2\,{\rm M_\odot\, yr^{-1}}$). Both galaxies are
    viewed along a line of sight parallel to the total stellar angular momentum vector. The circles in each panel highlight three
    measures of galaxy size used in this paper (see Section~\ref{SSSecSizes} for details): (1) the 2D half-stellar-mass radius,
    $R_{\star,50}$ (solid white circle); (2) the 2D half-baryonic-mass radius, $R_{\rm bar,50}$ (dashed light blue circle); and (3)
    the radius $R_1$ at which the stellar surface density profile first drops below $\Sigma_\star = 1\,{\rm M_\odot\,pc^{-2}}$
    (dot–dashed yellow circle). The image scale is indicated by the horizontal dashed line in the lower-right corner, which spans
    $15\,{\rm kpc}$.}
  \label{fig:examples}
\end{figure*}

Following \citet{Trujillo2020}, we also consider another size measure, the radius $R_1$ where the stellar surface density profile drops below
$\Sigma_1 \equiv 1\ \mathrm{M}_\odot\,\mathrm{pc}^{-2}$ (coloured squares in Fig.~\ref{fig:Sigma}). We calculate $R_1$
from the face-on projection of the stellar surface mass density profiles—defined in the plane perpendicular to the net angular momentum
vector of the bound stellar particles—by linearly interpolating between the radial bins that bracket $\Sigma_1$.
The $\Sigma_\star(R)$ profiles are constructed in the manner described above.
This mimics the observational analysis of \citet{Trujillo2020}. All bound stellar particles are used to calculate
$R_1$, regardless of their distance from the galaxy centre.\footnote{For the vast majority of galaxies, similar values
of $R_1$ are obtained when all stellar particles are used to construct their surface mass density profiles, regardless of whether or
not those particles are gravitationally bound to a galaxy.} For the lowest mass galaxies, $R_1$ typically exceeds
  $R_{\star,50}$ by factors of a few, increasing to $\approx 10$ for the most massive galaxies in our sample.

Finally, we examine the baryonic size--mass relation, which accounts for both stellar and gaseous components. It is defined by the half-baryonic-mass radius, $R_{{\rm bar},50}$, and the total
baryonic mass, $M_{\rm bar}$. Baryonic half-mass radii are derived from cumulative mass profiles (i.e. curves of growth, CoGs)
constructed by integrating the face-on stellar and gas surface-density profiles. 
For most galaxies, the gaseous component extends far into the halo and has no obvious ``edge''; the total baryonic masses and the corresponding
half-mass radii are thus poorly defined. To address this, we follow the procedure that \citet{Zichen2025} applied to SPARC galaxies and impose an
outer boundary at the radius $R_{\rm HI}$ where the \HI~column density first falls below $10^{20}\,{\rm cm^{-2}}$ (which is of order
$1\,{\rm M_\odot\,pc^{-2}}$). Gas surface-density profiles are computed out to this limiting radius using the \HI~masses of gas particles,
multiplied by a factor of 1.33 to account for helium and heavier elements.\footnote{This approach mimics the observational analysis of SPARC galaxies
by \citet{Zichen2025}, but neglects the contribution of ${\rm H_2}$ and ionised hydrogen in the galaxy outskirts. However, the
resulting baryonic mass–size relations obtained as described above are consistent with those obtained when including both ionised and molecular gas.}
Separate CoGs are constructed for the stellar and \HI~components and then combined to obtain $R_{\rm bar,50}$. The total baryonic
mass is the sum of the stellar mass, $M_\star$, and the \HI~mass (scaled by 1.33) enclosed within $R_{\rm HI}$.

To illustrate these size defintions, Fig.~\ref{fig:examples} shows two example galaxies from L025m5 as virtual \textit{Hubble Space Telescope}
(HST) observations. The left panel shows a star-forming disc galaxy ($M_\star=8.8\times 10^{10}\,{\rm M_\odot}$,
SFR=$2.5\,{\rm M_\odot\, yr^{-1}}$), while the right panel shows a passive spheroid ($M_\star=1.5\times 10^{11}\,{\rm M_\odot}$,
SFR=$0.2\,{\rm M_\odot\, yr^{-1}}$). In each panel, the circles indicate the projected stellar half-mass
radius, $R_{\star,50}$, the baryonic half-mass radius, $R_{\rm bar, 50}$, and the surface-density radius $R_1$, defined by
$\Sigma_\star(R_1) = 1 \, {\rm M_\odot \, pc^{-2}}$.

Both images were created using the algorithm described in \citetalias{Partridge}, which uses SPH interpolation to assign dust masses and
stellar luminosities to cells of a Cartesian grid centred on each galaxy. Stellar light is attenuated by dust, accounting for
both chemical composition and grain size using a 1D radiative transfer model (light scattered along the line of sight—a minor contribution
to the total flux—is neglected). The final images combine HST ACS/WFC\footnote{Hubble Space Telescope Advanced Camera for Surveys Wide
Field Channel.} F475W (blue), F625W (green), and F775W (red) filters.

To ensure that these size measurements are robust to numerical resolution, we restrict our analysis to central
galaxies with stellar masses $M_\star \geq 100\times m_{\rm bar}$ (see Appendix~\ref{app:convergence} for a
justification of this choice). For $R_{\star,50}^{\rm Ser}$ and $R_{r,50}^{\rm Ser}$, which are obtained from profile
fitting and are therefore more sensitive to particle noise, we impose a more conservative requirement of
$M_\star \geq 500\times m_{\rm bar}$.

\begin{figure*}
  \subfloat{\includegraphics[width=0.98\textwidth]{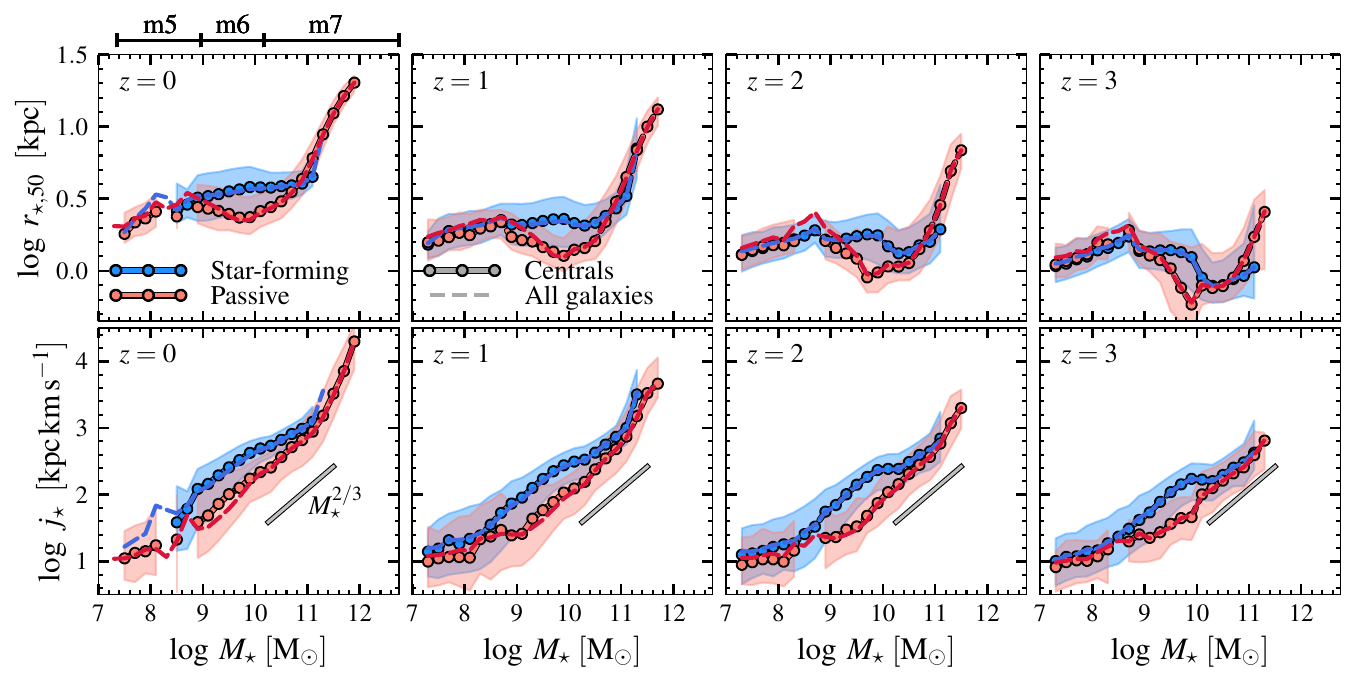}}
  \caption{Three-dimensional stellar half-mass radius ($r_{\star,50}$; upper panels) and stellar specific angular momentum ($j_\star$; lower panels)
    as a function of stellar mass at $z=0$, 1, 2, and 3 (left to right). Median relations are constructed by combining galaxies from different
    runs above a resolution-dependent mass limit. For m5 resolution (L025m5 at $z=0$, and L050m5 at $z>0$) we include all galaxies
    with $M_\star \ge 100\, m_{\rm bar}$; for L200m6 and L400m7, we include galaxies with  $M_\star \ge 500\, m_{\rm bar}$ and  $\geq 1000\, m_{\rm bar}$,
    respectively. The line segments above the upper-left panel indicate the stellar-mass range where each simulation contributes the highest number of
    galaxies. Connected circles show relations for central galaxies, while dashed lines show the entire galaxy population (centrals plus satellites).
    Shaded regions denote the $16^{\rm th}$–$84^{\rm th}$ percentile scatter for the full distribution of centrals.
    Blue lines show the median relations for galaxies within 0.5 dex of the redshift-dependent star-formation main sequence defined by eq.~\ref{eqSFR}
    (star-forming systems), while red lines show the medians for galaxies at least 0.5 dex below the main sequence (passive systems). 
    Only bins that contain at least 25 galaxies are plotted.
    The grey line segment in the lower panels shows the $j_\star\propto M_\star^{2/3}$ scaling predicted by the virial theorem.
    For $z\lesssim 2$ and $10^9\lesssim M_\star/{\rm M_\odot} \lesssim 10^{11}$, star-forming central galaxies are larger (by $\approx0.1$–0.3 dex) and have
    higher $j_\star$ (by $\approx0.2$–0.5 dex) than passive galaxies. 
    Conversely, massive ($M_\star \gtrsim 10^{11}\,{\rm M_\odot}$) and low-mass ($M_\star \lesssim 10^9\,{\rm M_\odot}$) star-forming and
    passive have similar sizes and angular momenta at all redshifts.
    Including satellite galaxies does not affect the trends, which can be seen by comparing the dashed lines to the connected circles.}
  \label{fig:Size_j_redshift}
\end{figure*}

\subsubsection{Galaxy angular momentum}
\label{SSSecAngMom}

The angular momentum of each galaxy is computed with respect to its most bound particle and centre-of-mass velocity of the relevant
component. Unlike our estimates of stellar mass and SFR, the stellar
particles used here are not restricted to any aperture; instead, we include all particles that are gravitationally bound to the
central galaxy.\footnote{Unlike our measurements of stellar mass and size, both of which are restricted to a ${\rm 50\,kpc}$
aperture, we do not impose an aperture when calculating $j_\star$, but instead use all bound stellar particles. This better reflects
observational attempts to estimate $j_\star$, which typically derive cumulative
$j_\star(<R)$ profiles out to large radii (often $\approx 5$--$10$ disc scale lengths) and often require convergence at large $R$.
When kinematic data or surface brightness profiles are limited to smaller radii, they are commonly extrapolated to estimate the total
angular momentum assuming a flat rotation curve. For late-type galaxies, the resulting $j_\star$ values are generally close to total specific angular
momentum of an exponential disc with scale length $R_d$ and flat rotation velocity $V_{\rm flat}$, for which $j_\star = 2\,R_d\,V_{\rm flat}$.

In simulations and observations, $j_\star$ is sensitive to aperture choice because stars at large radii contribute little
to the total mass but can carry substantial angular momentum. We quantify the impact of imposing a finite aperture in
Appendix~\ref{app:aperture}.}
We define the stellar specific angular momentum as
$j_\star = J_\star / M_{\star,{\rm bound}}$, where $J_\star$ is the magnitude of the angular momentum of bound stellar particles and
$M_{\star,{\rm bound}}$ is their total mass.

We also compute the baryonic angular momentum, $J_{\rm bar} = J_\star + J_{\rm gas}$, where $J_{\rm gas}$ is obtained
from gas particles within $R_{\rm HI}$, weighted by their \HI~fractions and multiplied by 1.33 to approximately
account for heavier elements (primarily helium). This follows the procedure used for SPARC galaxies
by \citet{Zichen2025}; we compare our results to theirs in Section~\ref{sec:JMB-lowz}. The specific baryonic angular momentum is then
$j_{\rm bar} = J_{\rm bar}/M_{\rm bar}$, with $M_{\rm bar} = M_{\star,{\rm bound}} + 1.33\,M_{\rm HI}$. Gas fractions are defined as
$f_{\rm gas}=1.33\times M_{\rm HI}/M_{\rm bar}$.

As with the baryonic half-mass radius, $J_{\rm bar}$ does not capture the full angular momentum of all baryonic material: it
excludes contributions from ionised and molecular gas, and treats heavy elements only approximately. We adopt this
simplified definition to enable a fair comparison with observational measurements of baryonic angular momentum by \citet{Zichen2025}.
A follow-up paper will present a more comprehensive analysis of the evolution of the total galactic baryonic angular momentum content
of \colibre~galaxies.

\subsection{Calibration versus prediction in simulation-observation comparisons}

\colibre~subgrid models for supernova and AGN feedback were calibrated to reproduce the median $z=0$ galaxy size--mass relation
of \citet{Hardwick2022} over the stellar mass range $10^9 < M_\star/{\rm M_\odot} < 10^{11.3}$. Their galaxy sample is drawn from
the xGASS survey \citep{Catinella2010,Catinella2018} and spans redshifts $0.01 < z < 0.05$.
Sizes were inferred from projected stellar mass surface density profiles. These were constructed from double-
and single-\citet{Sersic1963} profile fits to $g$-, $r$-, and $i$-band images by \citet{cook2019}, which were used to
independently model the disc and (if present) bulge components. The fitted profiles, obtained using ProFit \citep{Robotham2017},
were combined and converted to stellar mass surface density profiles using the stellar mass-to-light conversion of
\citet{Zibetti2009}. The surface mass density profiles were then integrated to $10\times R_{50,r}$ (where $R_{50,r}$ is the
$r$-band half-light radius from \citealt{cook2019}) to obtain the cumulative stellar mass; the total stellar mass corresponds
to the mass enclosed within $10\times R_{50,r}$, which is used to define the half-mass radius.

\colibre's calibration used two-dimensional stellar half-mass radii measured directly from the particle
data for galaxies of mixed morphology. Stellar masses and sizes were measured at $z=0$ within 50 kpc apertures and therefore
do not exactly match the observational techniques used to infer the galaxy sizes to which the subgrid models were calibrated.
Although obtaining a good match to these observations, as \colibre~does, is by no means trivial, comparisons of simulated and
observed size--mass relations based on
similar size definitions are not fully independent tests of the model, and some level of agreement is expected by construction if
the calibration was successful.

To provide a more independent test, the comparisons presented below employ datasets and measurement techniques that differ
from the size estimates used for calibration. These include $z=0$ size--mass relations based on alternative size definitions
(e.g. light-weighted sizes in multiple bands in Section~\ref{sec:SM}, sizes inferred from surface density thresholds in
Section~\ref{sec:SMR1}, or size estimates obtained from the entire baryonic content of galaxies in
Section~\ref{sec:SMRbar}), different measurement procedures (particle-based versus profile fitting; Section~\ref{sec:SM}),
or subsamples defined by morphology, star-formation activity, or gas fraction. These choices can introduce systematic
differences in the inferred galaxy sizes, so the level of agreement with observations depends on how robustly
\colibre~reproduces the internal structure of galaxies in the local Universe.

Truly predictive tests arise from observables that were not part of the calibration. These include the redshift evolution of
the size--mass relation (Section~\ref{sec:SM-highz}), relations obtained for independent samples of different galaxy types,
and all relations involving baryonic or stellar specific angular momentum (Section~\ref{sec:low-z-ang}). In particular,
the $z>0$ size--mass relations and all $j_\star$--$M_\star$ relations, as well as their dependence on galaxy type,
constitute genuine predictions of the model. The level of agreement with observations therefore provides a non-trivial
validation (or refutation) of the underlying physical models implemented in \colibre.

\begin{figure}
  \subfloat{\includegraphics[width=0.5\textwidth]{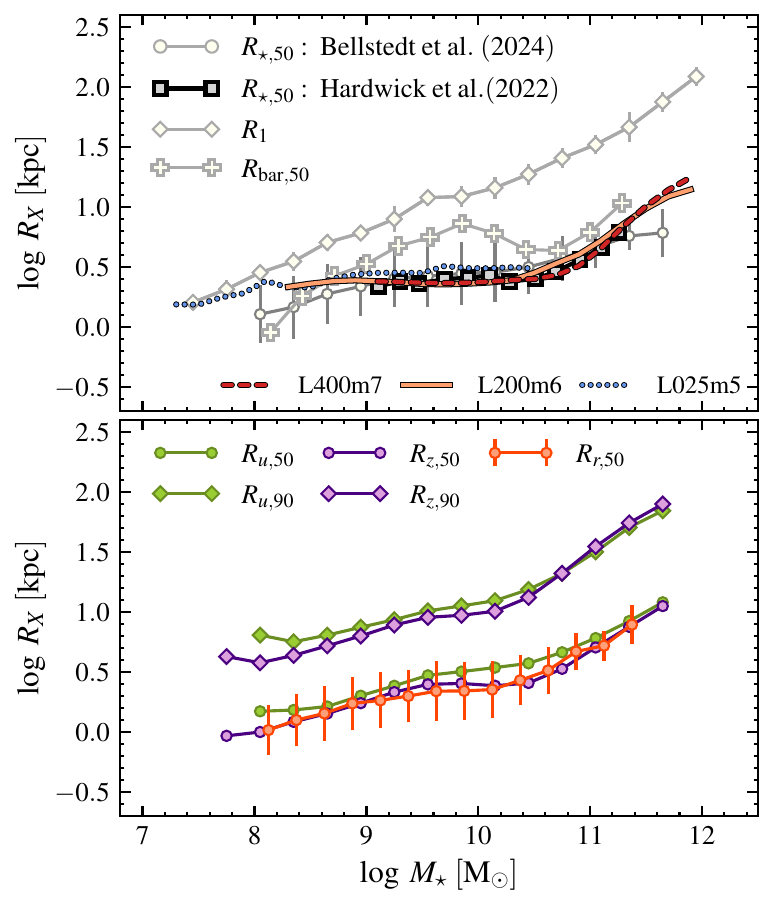}}
  \caption{The upper panel shows mass-weighted size estimates: the two-dimensional half-mass radii $R_{\star,50}$ from xGASS
    \citep{Hardwick2022} and GAMA \citep{Bellstedt2024}, the radius $R_1$ at which the stellar surface density equals
    $\Sigma_\star = 1\,{\rm M_\odot\,pc^{-2}}$ \citep{Trujillo2020}, and the baryonic half-mass radii of SPARC galaxies
    \citep{Zichen2025}. Also shown are the median $z=0$ two-dimensional half-mass radii of \colibre~galaxies in L025m5, L200m6,
    and L400m7 (coloured lines). The latter, shown for all galaxies (centrals plus satellites), were computed directly from the stellar particle data and were used to calibrate
    \colibre's sub-resolution physics models; these relations should only be compared with the observed $R_{\star,50}-M_\star$ relations. The lower panel shows
    light-weighted sizes from the GAMA Survey, including the $r$-band two-dimensional half-light radii $R_{r,50}$ \citep{Casura2022},
    and the radii enclosing 50 and 90 per cent of the $u$- and $z$-band light. Error bars, where present, correspond to
    the $16^{\rm th}$ to $84^{\rm th}$ percentile scatter.}
  \label{fig:Size_Observations}
\end{figure}

\begin{figure*}
  \subfloat{\includegraphics[width=0.85\textwidth]{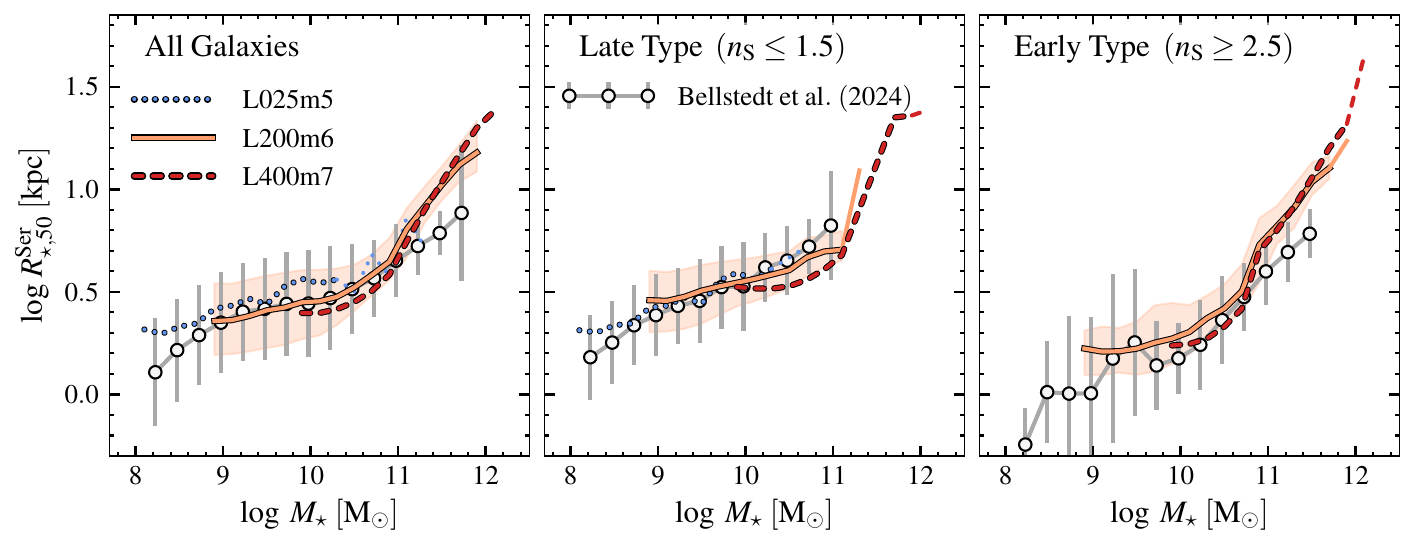}}\\\vspace{-0.5cm}
  \subfloat{\includegraphics[width=0.85\textwidth]{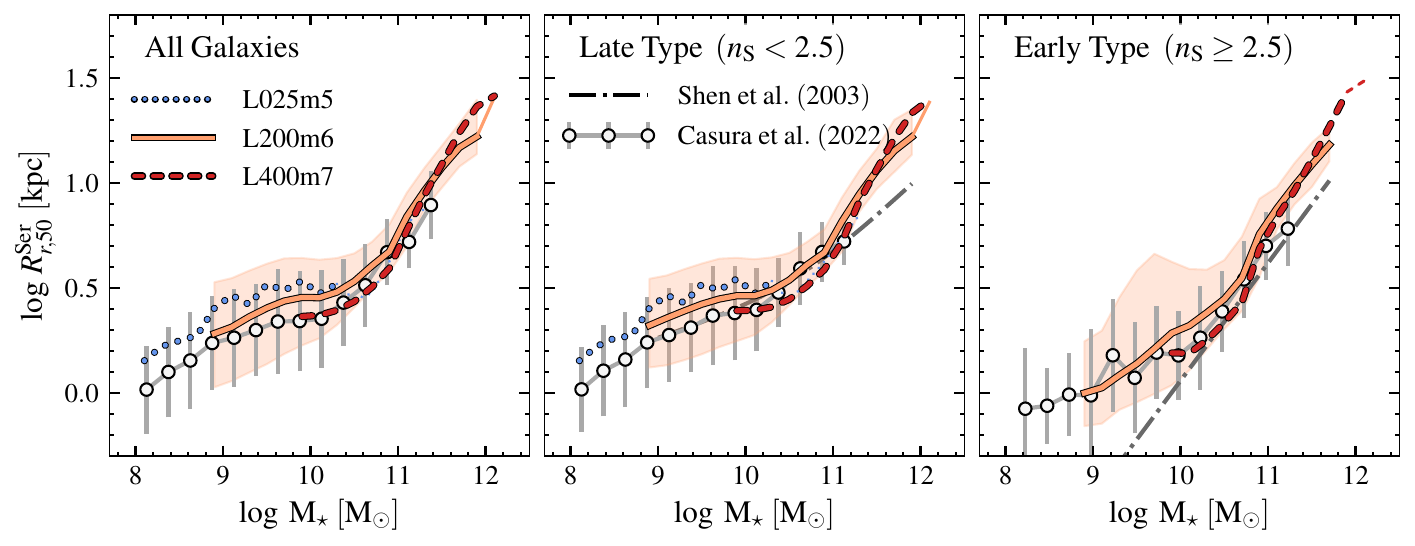}}
  \caption{Size–mass relations for $z=0$ central galaxies in the \colibre~simulations compared with recent observations.
    The top panels show median 2D stellar half-mass radii, $R^{\rm Ser}_{\star,50}$, compared with observational measurements from \citet{Bellstedt2024}.
    The bottom panels show the median 2D $r$-band half-light radii, $R^{\rm Ser}_{r,50}$, and are compared with the measurements of \citet{Casura2022}
    (connected circles) and \citet{Shen2003} (dot-dashed lines).
    In both cases, galaxy sizes were measured from single-component Sérsic fits to the stellar mass surface density or $r$-band surface
    brightness profiles, as described in Section~\ref{SSecAnalysis}. Results from L025m5, L200m6, and L400m7 are shown using dot-dashed,
    solid, and dashed lines, respectively, with thin line segments indicating mass bins that contain fewer than 10 galaxies. 
    The shaded regions highlight the $16^{\rm th}$ to $84^{\rm th}$ percentile scatter for L200m6. Observed data are plotted as connected
    symbols, with error bars indicating the $16^{\rm th}$–$84^{\rm th}$ percentile range. The left column shows results for all central
    galaxies, regardless of morphological type. The middle and right columns show late- and early-type central galaxies, respectively,
    for which morphological classification is based on Sérsic
    index, $n_{\rm S}$: in the top (bottom) panels, galaxies with $n_{\mathrm{S}} \leq 1.5$ (2.5) are classified as late-type, and those with
    $n_{\mathrm{S}} \geq 2.5$ as early-type. Overall, \colibre~reproduces the size--mass relations for both early- and late-type galaxies, and
    achieves good convergence across resolution levels.} 
  \label{fig:Size_Mass_Casura}
\end{figure*}

\subsection{An initial view of the $r_{\star,50} - M_\star$ and $j_\star - M_\star$ relations}
\label{SSecr50jstar}

Fig. \ref{fig:Size_j_redshift} shows the median $r_{\star,50} - M_\star$ (upper panels) and 
$j_\star - M_\star$ relations (lower panels) for central galaxies in L025m5 (or L050m5 for available
redshifts), L200m6, and L400m7 using connected circles. Dashed lines show relations for the full galaxy population,
including satellites. For clarity, galaxies from all simulations have been combined to construct a single
median relation. To balance the requirements of high resolution and good statistics, we adopt a resolution-dependent
lower mass limit corresponding to $100\, m_{\rm bar}$, $500\, m_{\rm bar}$, and $1000\, m_{\rm bar}$ for the m5, m6, and m7 resolutions,
respectively. 
The line segments above the upper-left panel indicate the range of stellar mass
over which each resolution level contributes the largest number of galaxies. Convergence with resolution is
demonstrated explicitly in later figures. 

Different columns correspond to results at $z=0$, 1, 2, and 3. Blue curves show the median relations for star-forming galaxies,
i.e. those lying within 0.5 dex of the redshift-dependent star formation main sequence,\footnote{Equation~\ref{eqSFR} provides a useful proxy
for the star-forming main sequence. It was obtained by fitting a second-order polynomial to the combined median ${\rm sSFR}$–$M_\star$ relations
calculated independently for L025m5 (for $z\leq 0.5$), L050m5 (for $z\geq 1$), L200m6, and L400m7, at redshifts $z=0$, 0.5, 1, 2, 3, and 4,
and over the stellar-mass range $7.5 \leq \log(M_\star/{\rm M_\odot}) \leq 10$ in bins of width $\Delta\log M_\star = 0.2$.} approximated  as
\begin{equation}
  \log\, [{\rm sSFR}(z)/{\rm Gyr}^{-1}] = 0.40\, z - 0.04\, z^2 - 0.73
  \label{eqSFR}
\end{equation}
Red curves show medians for passive galaxies, i.e. those that lie at least 0.5 dex below this sequence. We choose these thresholds in order
to sample the full population of galaxies at all redshifts. In subsequent figures, we adopt definitions of star-forming and passive galaxies
that are tailored to the observational datasets against which \colibre~is compared.

Key features of the size–mass relations in Fig.~\ref{fig:Size_j_redshift} are as follows. (i) The median $r_{\star,50} - M_\star$ relations for
both galaxy types are approximately flat below a characteristic mass of $M_\star \approx 10^{10.5}\,{\rm M_\odot}$, although star-forming galaxies at $z\leq 1$
exhibit a slight increase in size with stellar mass.
(ii) At all masses, galaxy sizes increase towards lower redshift, irrespective of galaxy type. 
(iii) At intermediate masses ($10^9 \lesssim M_\star/{\rm M_\odot} \lesssim 10^{11}$ at $z\leq 2$ and $10^9 \lesssim M_\star/{\rm M_\odot} \lesssim 10^{10}$
at $z\geq 2$ ), star-forming galaxies are, on average,
larger than passive systems by $\approx 0.2$–0.3 dex, regardless of whether satellite galaxies are included.
(iv) Among the most massive systems, sizes
increase sharply with mass at low redshift, but by $z=3$ the relation develops a pronounced `U'-shape, with sizes decreasing to a
minimum near $M_\star \approx 10^{10.5}\,{\rm M_\odot}$ before rising rapidly at higher masses. (v) At $z=3$, this “compaction’’ mass scale roughly
separates galaxies with the highest sSFRs ($\log[{\rm sSFR}/{\rm Gyr}^{-1}] \gtrsim 0.5$; typically less massive) from quenched systems
($\log[{\rm sSFR}/{\rm Gyr}^{-1}] \lesssim -2$; typically more massive), although this is not shown explicitly in Fig. \ref{fig:Size_j_redshift}.

Key features of the $j_\star$--$M_\star$ relations are as follows. (i) At intermediate mass and for all redshifts, star-forming galaxies have higher median
$j_\star$ than passive ones by $\approx 0.2$ to 0.5 dex, depending on mass.
(ii) The $j_\star$--$M_\star$ relations of both star-forming
and passive galaxies follow approximate power laws with slopes $\approx 0.7$ (cf.\ the solid grey line).
(iii) At the highest masses and for $z \leq 1$, there is evidence for an upturn in $j_\star$ for passive galaxies, though
this trend is sensitive to the aperture used to measure $j_\star$ (see Appendix~\ref{app:aperture}). (iv) At all redshifts, active
and passive galaxies with $M_\star \approx 10^{11}\,{\rm M_\odot}$ have similar $j_\star$ values regardless of their sSFR.
These results are unaffected by the exclusion of satellite galaxies.

\begin{figure}
  \subfloat{\includegraphics[width=0.5\textwidth]{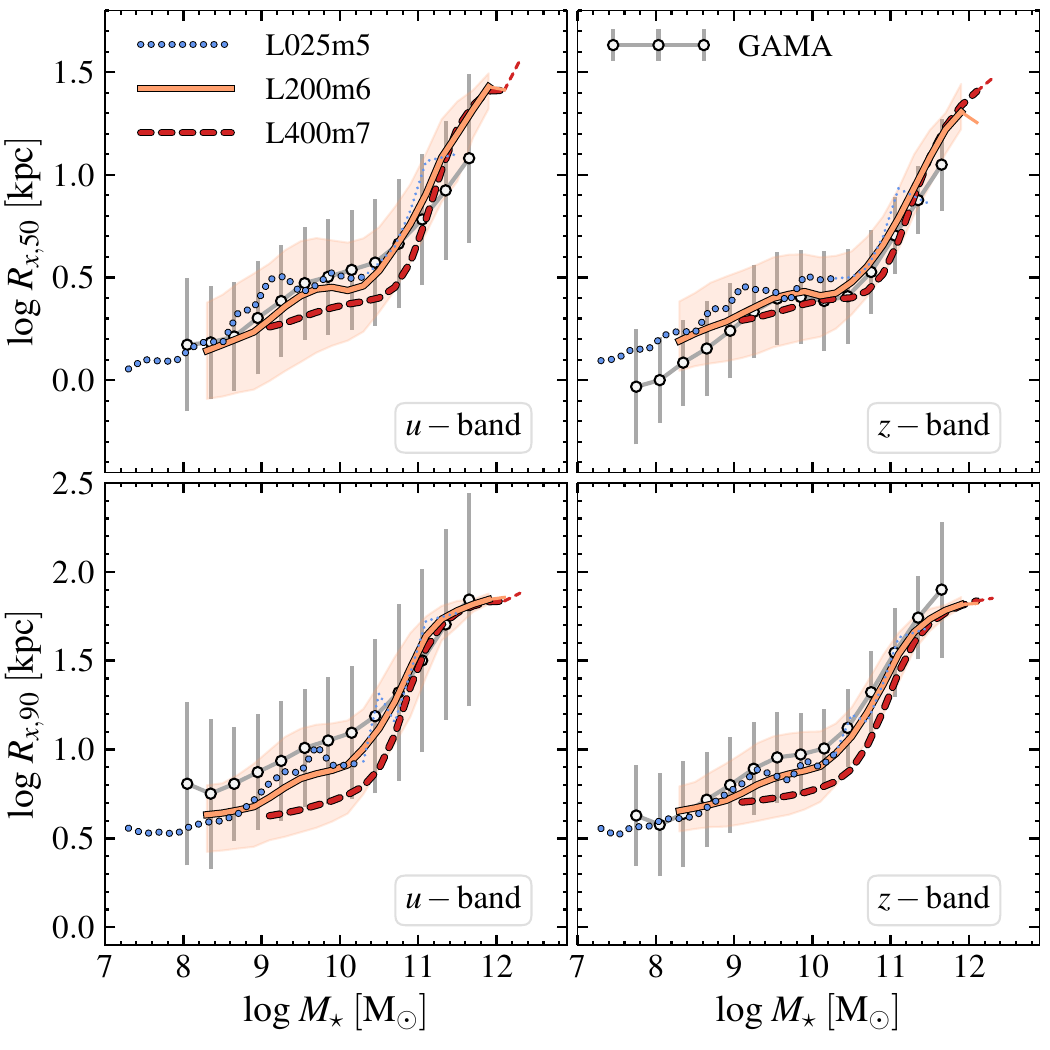}}
  \caption{Characteristic radii $R_{x,50}$ (upper panels) and $R_{x,90}$ (lower panels) where ``$x$'' represents the $u$-band (left column)
    and $z$-band (right column) as a function of stellar mass. Coloured lines show the median relations for simulated central galaxies at $z=0$
    and do not include the effects of dust attenuation;
    thick and thin lines correspond to bins that contain $\geq 10$ and $< 10$ galaxies, respectively.
    Shaded regions denote the $16^{\rm th}$ to $84^{\rm th}$ percentile scatter for L200m6. Connected circles show
    the median observational relations from the GAMA survey, with error bars indicating the $16^{\rm th}-84^{\rm th}$ percentile range.}
  \label{fig:Size_GAMA_u_z}
\end{figure}

\section{Results}
\label{ref:results}

\begin{figure}
  \subfloat{\includegraphics[width=0.5\textwidth]{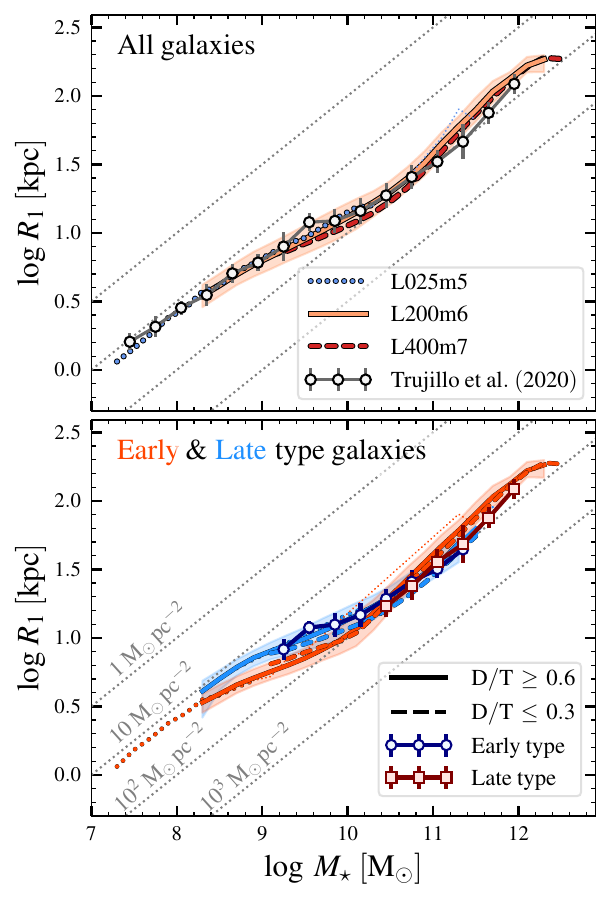}}
  \caption{Size–mass relations for $z=0$ central galaxies compared with the
    observational results of \citet{Trujillo2020}. Sizes are defined by the radius $R_1$ where the (face-on) stellar
    surface mass density drops below $\Sigma_1 = 1\,\mathrm{M}_\odot\,\mathrm{pc}^{-2}$.
    Different line styles correspond to median relations for different resolution levels. Shaded
    regions and error bars show the $16^{\rm th}$ to $84^{\rm th}$ percentile range for the (L200m6) simulations and observations, respectively.
    The upper panel shows the full samples of observed and simulated galaxies; the lower panel separates
    galaxies by morphological type. Observed morphologies (available for galaxies with $M_\star \gtrsim 10^9\,{\rm M}_\odot$)
    are based on visual classifications: ellipticals (E0–S0+) are shown in red, spirals (S0/a–Im) in blue. For \colibre,
    we use the kinematic disc-to-total ratio, $D/T$, as a proxy for morphology, as indicated in the legend. The grey dotted lines correspond to
    tracks of constant stellar surface density. \colibre~demonstrates excellent convergence across resolution levels and aligns closely with observed data.}
  \label{fig:Size_Mass_Trujillo}
\end{figure}

\subsection{Size--mass relations at $z=0$}
\label{sec:SM-lowz} 

In this section, we compare the size--mass relations of \colibre~galaxies with observational datasets that were
not used to calibrate the sub-grid models. Although each survey employs its own methodology to estimate stellar
masses and sizes, several common elements can be summarised here.

Stellar masses are typically derived from multi-band optical imaging combined with stellar population synthesis
modelling, with corrections for dust and AGN and an assumed initial mass function. For a fixed initial mass function,
this results in random uncertainties
of $\approx 0.1$--$0.3$ dex in $M_\star$, depending on redshift and data quality in addition to unknown systematic uncertainties
\citep[e.g.][]{Gallazzi2009,Conroy2009b,Robotham2020,Pacifici2023}. When comparing to observational results, we
therefore apply a redshift-dependent lognormal scatter to the simulated stellar masses to mimic these
uncertainties.\footnote{\label{fn:obserr}Following \citet{Chaikin2026b}, we add a lognormal scatter to $M_\star$ with mean zero and
standard deviation $\sigma_{\rm random}(z)=\min(0.1+0.1\,z,\,0.3)\,{\rm dex}$, consistent with the functional form
advocated by \citet{Behroozi2010}, although with slightly different parameters. Specific star formation rates and specific
angular momenta are calculated using the measured (i.e. unscattered) values of $M_\star$.
We examine the impact of redshift-dependent scatter in $M_\star$ on the SMR and AMR in Appendix~\ref{app:scatter}.}

The datasets considered below span a range of depths and modelling approaches. The $z=0$ SMR of \citet{Hardwick2022}
is based on xGASS galaxies whose structural parameters were derived using \textsc{profit}
\citep{Robotham2017}. For GAMA galaxies, stellar masses, surface-density (and surface-brightness)
profiles, and sizes were obtained from updated multi-band photometry using the
\textsc{profuse} pipeline \citep{Robotham2022}, adopting the stellar population synthesis models of
\citet{BruzualCharlot2003}, a \citet{Chabrier2003b} initial mass function, and a \citet{Calzetti2000} dust law.
The IAC Stripe82 Legacy Project \citep{FliriTrujillo2016,RomanTrujillo2018}, analysed by \citet{Trujillo2020},
provides deep multi-band imaging corrected for inclination and dust attenuation; stellar masses are derived by
integrating the surface mass-density profiles to an isophotal limit of $g=29\,\mathrm{mag\,arcsec^{-2}}$. The SPARC
survey comprises 175 nearby late-type galaxies whose stellar mass distributions are inferred from 3.6$\mu$m surface
photometry obtained with the \textit{Spitzer Space Telescope}.

We compare these datasets in Fig.~\ref{fig:Size_Observations}, which summarises the range of observationally inferred $z=0$ size--mass
relations considered in this work. The upper panel shows mass-weighted size estimates, including the two-dimensional half-mass radii
$R_{\star,50}$ of GAMA galaxies from \citet{Bellstedt2024}, the surface-density threshold radius $R_1$ of \citet{Trujillo2020}, and the
baryonic half-mass radii $R_{\rm bar,50}$ of \citet{Zichen2025}. Also shown is the $R_{\star,50}$--$M_\star$ relation of \citet{Hardwick2022},
which was used to calibrate \colibre's subgrid physics models. For reference, we include the corresponding $z=0$ $R_{\star,50}$--$M_\star$ relations for
all galaxies in the L025m5, L200m6, and L400m7 simulations, where $R_{\star,50}$ is measured directly from random projections of the
stellar particle distributions. These simulated relations show good convergence and are in excellent agreement with the SMR of
Hardwick et al. The simulated $R_{\star,50}$ values were used to calibrate \colibre's subgrid physics models and should be compared only to this
dataset; they are shown for completeness but
are not analysed further, as they have been presented previously by \citet{Schaye2026} and \citet{Chaikin2026a}. The different size definitions
exhibit substantial systematic differences, and we therefore defer detailed comparisons with the corresponding observational datasets to subsequent
sections. 

The lower panel shows light-weighted size estimates from GAMA Data Release
4,\footnote{\href{https://www.gama-survey.org/dr4/}{https://www.gama-survey.org/dr4/}.} including the
half-light radii in the $r$-, $z$-, and $u$-bands, and the radii $R_{u,90}$ and $R_{z,90}$ enclosing
90 per cent of the $u$- and $z$-band light, respectively. 

In the following sections, we compare these observational relations to \colibre's simulated galaxy population. For more rigorous tests
of \colibre's predictions, we will sometimes present results separately for disc and spheroid subsamples, or for star-forming
and passive systems, using similar selection criteria for the simulated and observed galaxies.
Hereafter, we restrict our analysis to central galaxies in the thermal-AGN model. For the observational datasets
we do not distinguish between central and satellite galaxies; however, we demonstrate in Appendix~\ref{app:centsat} that our
$z=0$ results are insensitive to the removal of satellites from the simulated sample.

In all subsequent figures, simulated and observed size--mass and angular momentum--mass relations are constructed using bins of
fixed logarithmic width, $\Delta\log M_\star=0.2$. The plots follow a common
visual scheme: connected symbols with error bars denote observed median relations and their $16^{\rm th}$ to $84^{\rm th}$
percentile scatter, while lines and shaded regions show the corresponding medians and scatter from the simulations. When
galaxies are separated by morphological type, late-types are shown in blue and early-types in red; different resolution levels
are distinguished by line style. When all galaxies are shown irrespective of type, simulations are distinguished by line
style and colour. Across all figures, a given resolution level is represented by a consistent line style.

\begin{figure}
  \subfloat{\includegraphics[width=0.48\textwidth]{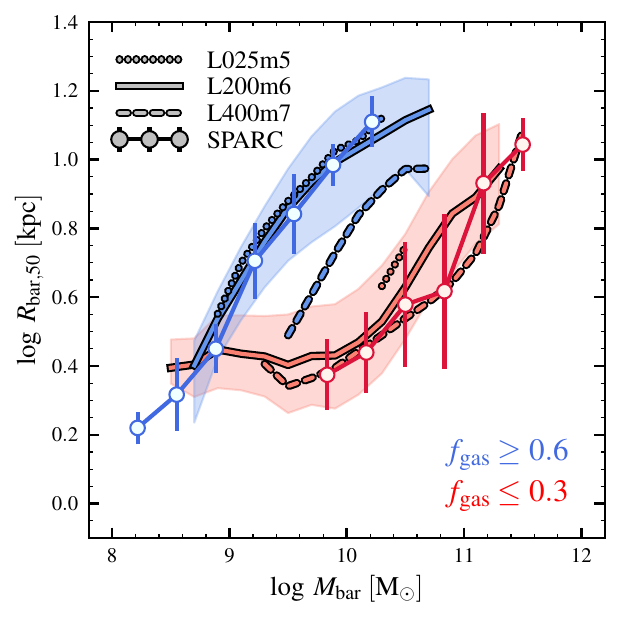}}
  \caption{The $z=0$ baryonic SMR for late-type (i.e. kinematic $D/T \geq 0.6$) central \colibre~galaxies, compared with disc galaxies
    from the SPARC survey \citep{Zichen2025}. Dotted, solid, and dashed curves show the median relations for L025m5, L200m6, and L400m7,
    respectively. Each simulation sample is divided into gas‑rich ($f_{\rm gas} \geq 0.6$; blue) and gas‑poor ($f_{\rm gas} \leq 0.3$;
    red) subsamples, where $f_{\rm gas} = 1.33\,M_{\rm HI}/M_{\rm bar}$. Connected circles denote the corresponding SPARC relations,
    selected with the same gas-fraction thresholds and plotted using the same colour scheme. Both simulated and observed relations show strong segregation by
    gas fraction. Overall, \colibre~reproduces the SPARC relations but the lower-resolution run (L400m7) contains an excess of overly compact gas-rich systems.}
  \label{fig:Rbar_Mstar}
\end{figure}

\subsubsection{Stellar size--mass relations}
\label{sec:SM}

The upper panels of Fig.~\ref{fig:Size_Mass_Casura} show the median $R^{\rm Ser}_{\star,50}$--$M_\star$ relations obtained from single-component
\sersic~fits to the stellar surface-density profiles, $\Sigma_\star(r)$, of $z=0$ central galaxies containing at least
500 stellar particles. The left panel shows results for the full sample, regardless of morphology. The middle and
right panels separate galaxies by best-fitting Sérsic index, with $n_{\rm S} \leq 1.5$ for late-types (middle) and $n_{\rm S} \geq 2.5$
for early-types (right). Different resolution runs are plotted individually and demonstrate good convergence. Thick
lines denote mass bins containing more than ten galaxies, while thin lines indicate bins with fewer; shaded regions 
indicate the scatter in the L200m6 run. In \colibre, the median $R^{\rm Ser}_{\star,50}$ is relatively flat below
$M_\star \approx 10^{10.5}\,\mathrm{M_\odot}$, but increases by roughly $1$ dex over the subsequent $\approx 1.5$ dex increase
in stellar mass. 

Circles with error bars show the observational relations for GAMA galaxies from \citet{Bellstedt2024}, who used updated multi-band
photometry and the PROFUSE \citep{Robotham2022} pipeline to the construct stellar mass maps that were used to infer the surface
mass density profiles. As for the simulations, structural parameters for the observed sample are obtained from single-component
Sérsic fits, with galaxies classified as pure discs ($n_{\rm S} \leq 1.5$; middle panel) or pure spheroids ($n_{\rm S} \geq 2.5$;
righthand panels). Error bars indicate the $16^{\rm th}$ to $84^{\rm th}$ scatter.

The lower panels show the $r$-band SMRs, compared to the measurements of \citet{Casura2022}, derived from
KiDS $r$-band imaging of the GAMA-II equatorial region \citep{Driver2011,deJong2013}. Sizes were obtained
from single-component Sérsic fits using \textsc{profit} \citep{Robotham2017}. The left panel shows the full population,
while the middle and right panels present the late- and early-type subsamples, adopting the same $n_{\rm S}$ thresholds for the
simulations and observations. For comparison, these panels also include the SDSS $r$-band power--law size--mass relations of
\citet{Shen2003} as dot-dashed lines.

Overall, we find excellent agreement between the simulated and observed measurements of $R^{\rm Ser}_{\star,50}$ and $R^{\rm Ser}_{r,50}$ across
the entire resolved stellar mass range. We have also compared size estimates in other optical bands, again obtaining good
consistency. Fig.~\ref{fig:Size_GAMA_u_z} illustrates this using the $R_{u,50}$--$M_\star$ (upper left) and
$R_{z,50}$--$M_\star$ (upper right) relations, where sizes were measured directly from the particle data rather than from S\'{e}rsic fits.
Grey points show the corresponding GAMA measurements derived in the same manner as $R_{r,50}$ from \citet{Casura2022}. Although our measurements do
not account for dust attenuation, we nevertheless find good agreement in both optical bands
(which are separated by roughly $250$–$300\,\mathrm{nm}$) in both the normalisation of the relations and their intrinsic scatter.

The lower panels of Fig.~\ref{fig:Size_GAMA_u_z} show that this agreement extends to $R_{u,90}$ and $R_{z,90}$, which
probe larger radii that were not considered during calibration of \colibre's sub-grid physics models. The
agreement is particularly good in the $z$-band. However, for simulated galaxies with $M_\star \lesssim 10^{10.5}\,{\rm M_\odot}$,
the $R_{u,90}$ values are on average slightly smaller than observed (but still within the scatter), possibly reflecting the
stronger sensitivity of shorter-wavelength light to dust, the effects of which are not included in our analysis of
simulated galaxy sizes.

\begin{figure*}
  \subfloat{\includegraphics[width=0.9\textwidth]{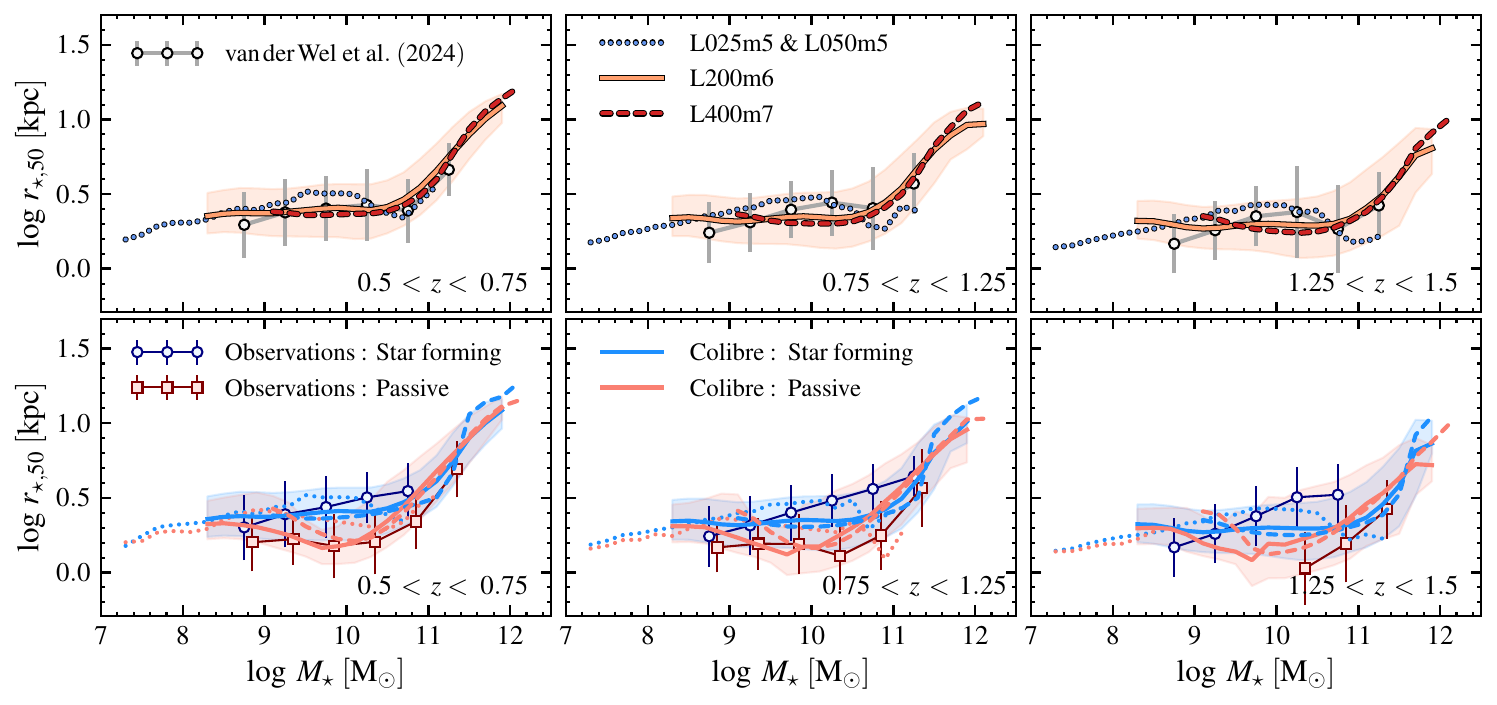}}
  \caption{Median $r_{\star,50}$–$M_\star$ relations at redshifts $0.5 < z < 1.5$.
    The upper panels show all observed galaxies and simulated central galaxies; the lower panels show median trends for
    subsets of star-forming (blue) and passive (red) galaxies separately. These classifications are based on each galaxy's distance
    from the $z$-dependent star formation main sequence (see text for details). Median relations for m5 (which combine L025m5 and L050m5),
    L200m6, and L400m7 are shown as dotted, solid, and dashed lines, respectively, to illustrate numerical convergence. In all panels,
    shaded regions indicate the $16^{\rm th}$ to $84^{\rm th}$ percentile scatter for L200m6. Simulated data include all
    available outputs within the redshift intervals listed in each panel. 
    Observational results from \citet{vanderWel2024} are shown for comparison.
    Their medians are shown with connected circles for the full sample in the upper panels and for the star-forming sample in
    the lower panels; connected squares in the lower panels show medians for the passive sample. Observational error bars
    denote the $16^{\rm th}$ to $84^{\rm th}$ percentile range. The simulations reproduce the observed redshift-dependence of the size--mass
    relation, as well as its dependence on star formation activity for $z\lesssim 0.75$. However, for galaxies with
    $M_\star \gtrsim 10^{10}\,{\rm M}_\odot$ at $z\gtrsim 0.75$, the separation between the active and passive
    relations is smaller than observed.}
  \label{fig:Size_Mass_vdWel}
\end{figure*}

\subsubsection{The $R_1 - M_\star$ relation}
\label{sec:SMR1}
Fig. \ref{fig:Size_Mass_Trujillo} compares the $R_1 - M_\star$ relation for $z=0$ central galaxies in \colibre~with
observations from \citet{Trujillo2020}. The top panel shows the full sample of observed and simulated galaxies. The bottom panel
splits the observed galaxies into discs and spheroids based on visual classifications.\footnote{\citet{Trujillo2020} use deep imaging
data from the IAC Stripe82 Legacy Project \citep{Fliri2016,Roman2018}, and target galaxies have redshifts $z< 0.09$ and span the mass range
$10^7\,\msun < M_\star < 10^{12}\,\msun$. Galaxies with $M_\star >10^9\,\msun$ were taken from the catalogue of \citet{Nair2010},
which includes visually classified morphologies. These data were supplemented with lower mass galaxies
($10^7\,\msun < M_\star < 3\times 10^9\,\msun$) from the catalogue of \citet{Maraston2013}, which lacks morphological information.}
\colibre~galaxies lack visual classification and are divided by the kinematic $D/T$ ratio: $D/T \geq 0.6$ for
late-types and $D/T \leq 0.3$ for early-types. The values of $R_1$ obtained for our simulated galaxies are
converged and agree well with the observational results over $\approx 5$ dex in stellar mass
\citep[see also][for similar results from other cosmological simulations]{ArjonaGalvez2025,DallaVecchia2025}.
Low-mass galaxies ($M_\star\lesssim 10^{8.5}\,{\rm M_\odot}$)
exhibit a roughly constant stellar surface density of $\Sigma_\star\approx 10\,{\rm M_\odot\, pc^{-2}}$, which increases to 
$\Sigma_\star\approx 100\,{\rm M_\odot\, pc^{-2}}$ for more massive systems ($M_\star\gtrsim 10^{10.5}\,{\rm M_\odot}$).
\colibre~also approximately reproduces the morphology dependence of the $R_1 - M_\star$
relation seen in the lower panel, which is weak for both the simulations and observations.

\subsubsection{The baryonic size--mass relation}
\label{sec:SMRbar}

Finally, Fig.~\ref{fig:Rbar_Mstar} compares the $z=0$ baryonic SMR of late‑type central \colibre\ galaxies (defined as kinematic
$D/T \geq 0.6$) with observations of late-type galaxies from the SPARC survey \citep{Zichen2025}. Dotted, solid, and dashed curves show the
median relations for L025m5, L200m6, and L400m7, respectively. Each sample is split into gas‑rich ($f_{\rm gas}\geq 0.6$; blue) and
gas-poor ($f_{\rm gas}\leq 0.3$; red) subsamples, where $f_{\rm gas} = 1.33\,M_{\rm HI}/M_{\rm bar}$. Connected squares indicate
the corresponding relations for SPARC galaxies, selected using the same gas‑fraction thresholds and plotted with the same colour scheme.
Both the simulated and observed relations show strong segregation by gas fraction.

There is good agreement between the median relations from L025m5 and L200m6 for the gas-rich samples, and between all
simulations for the gas-poor samples, with all matching the observational relations within the observed scatter. This agreement is
noteworthy given the heterogeneous selection function of the SPARC sample. The median relation from L400m7 for the gas-rich subsample,
however, is offset to smaller $R_{\rm bar,50}$ at fixed baryonic mass, suggesting the presence of a population of compact gas-rich
galaxies in this low-resolution run.

\vspace{0.5cm}
Overall, these results demonstrate that \colibre~accurately reproduces the characteristic galaxy sizes at $z=0$ inferred from optical
surface-brightness profiles, stellar surface-density profiles, and the low-density outskirts of galaxies quantified by the radius
$R_1$. This consistency across multiple, physically distinct size definitions provides confidence that at $z=0$ the spatial distribution
of both stars and baryons in \colibre~galaxies is realistic.

\begin{figure*}
  \subfloat{\includegraphics[width=1.\textwidth]{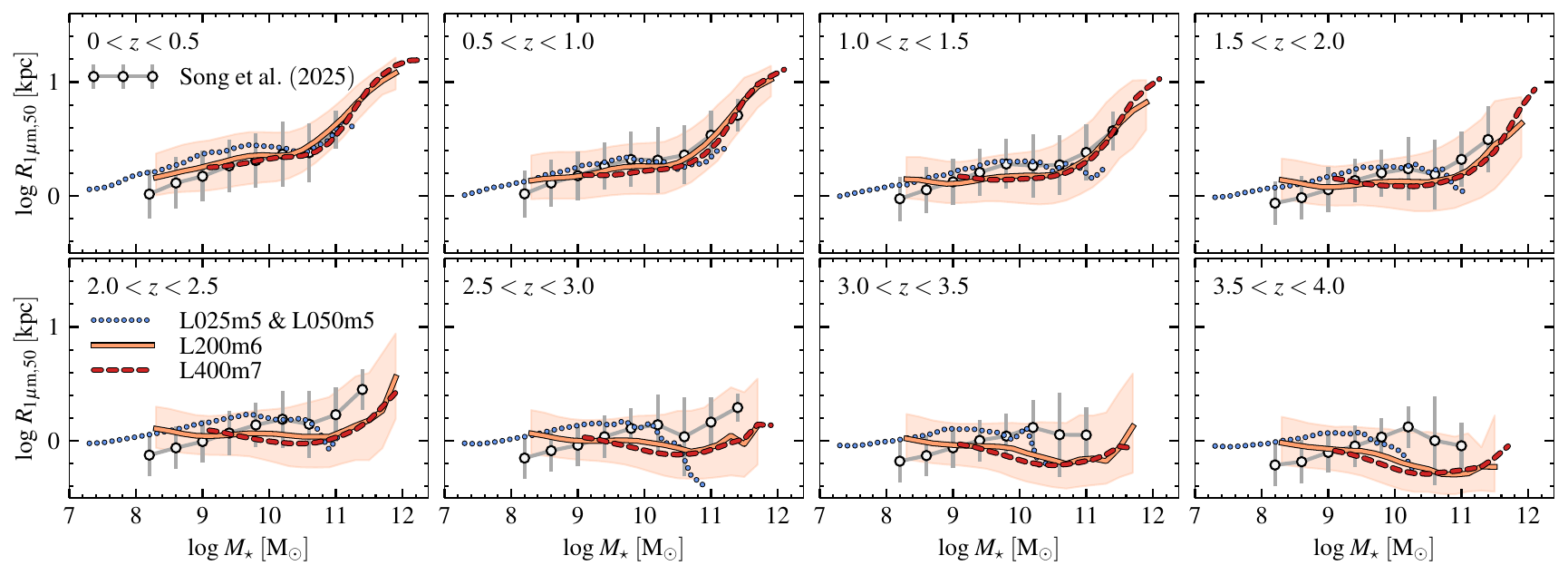}}
  \caption{Two-dimensional half-light radii, $R_{1\mu{\rm m},50}$, as a function of stellar mass,
    $M_\star$, for all observed and simulated central galaxies over the redshift range $0 \leq z \leq 4$. Sizes are estimated from $1\mu{\rm m}$
    surface brightness profiles measured within 50 kpc apertures and for random orientations. Different panels correspond to
    redshift bins of width $\Delta z=0.5$. Dotted, solid, and dashed coloured lines show the median relations for m5 (combining all available
    outputs from L025m5 and L050m5), L200m6, and L400m7, respectively; the orange shaded region shows the $16^{\rm th}$ to $84^{\rm th}$
    percentile scatter for L200m6. Median relations for observed galaxies from \citet{Song2026} are show using connected circles;
    error bars indicate the $16^{\rm th}$–$84^{\rm th}$ percentile range. 
    Observed sizes correspond to half-light radii measured from rest-frame $1\,\mu{\rm m}$ images. We approximate these measurements
    in \colibre~using half-light radii derived for the GAMA $Y$-band ($1.02\,\mu{\rm m}$).}
  \label{fig:Size_Mass_Song}
\end{figure*}

\subsection{Redshift evolution of the stellar size--mass relation}
\label{sec:SM-highz}

For higher-redshifts, we compare \colibre~with the recent measurements of \citet{vanderWel2024}, based on multi-band
HST imaging. Stellar masses and star-formation rates are from the \citet{Leja2020} catalogue, which used the 
Prospector-$\alpha$ model \citep{Leja2017, Johnson2021} 
to perform non-parametric modelling that accounts for dust attenuation and obscured AGN.
We also compare our simulated galaxies to recent observational results of \citet{Song2026}, based on HST + JWST multi-band imaging.
Stellar masses for this dataset were obtained from dust-corrected multi-band images using \textsc{CIGALE} \citep{Boquien2019}.

Fig. \ref{fig:Size_Mass_vdWel} shows the (3D) stellar half-mass radius, $r_{\star,50}$, as a function of stellar mass for
central \colibre~galaxies across several redshift intervals. The upper panels present results for central galaxies in the m5
resolution runs (combining L025m5 and L050m5; dotted lines), L200m6 (solid lines), and L400m7 (dashed lines). We include measurements
from all available snapshots that span each of the redshift intervals. Connected grey circles represent observed galaxies from
\citet{vanderWel2024}. Their 3D half-mass radii were derived by de-projecting 2D half-mass radii, $R_{\star,50}$, measured from
stellar mass maps derived from multi-band CANDELS HST imaging. 

The lower panels of Fig.~\ref{fig:Size_Mass_vdWel} separate galaxies into star-forming and passive subsamples based on their star formation
rates, using the same classification criteria for the simulated and observed galaxies. Galaxies are classified as passive if their
star formation rates place them at least 0.8~dex below the star-forming main sequence at their respective redshifts; otherwise, they are
classified as star-forming. This scheme is the same as that adopted by \citet{vanderWel2024}, who define the main sequence using the
parameterisation of \citet{Leja2022}. For \colibre~galaxies, we approximate the redshift-dependent star-forming main sequence 
using eq.~\ref{eqSFR}.

Overall, \colibre~accurately reproduces the redshift evolution of the observed stellar mass--size relation at $z \lesssim 1.5$,
including the qualitative differences between star-forming and passive populations. The agreement is particularly good at $z \lesssim 0.75$,
and remains so at higher redshifts for galaxies with $M_\star \lesssim 10^{10}\,{\rm M_\odot}$. For higher stellar masses,
systematic offsets are apparent at $z\gtrsim 0.75$: passive galaxies in \colibre~tend to be larger than observed by
$\approx 0.1$~dex, while star-forming galaxies are smaller by $\approx 0.2$~dex. As a result, the separation of the SMRs for star-forming
and passive galaxies in \colibre~is smaller than observed. These differences---particularly for star-forming galaxies---may be reduced once
  the effects of dust attenuation on inferred galaxy sizes are explicitly modelled, as dust can preferentially obscure the central regions of galaxies
  and increase measured sizes. Note, however, that star-forming galaxies in the m5
runs approximately reproduce the observed relations at all redshifts.
  
Fig.~\ref{fig:Size_Mass_Song} extends the comparison up to $z=4$. Observational measurements are taken from \citet{Song2026},
who combined multi-band photometry to derive galaxy sizes corresponding to rest-frame $1\,\mu{\rm m}$ imaging. These wavelengths are
expected to be less sensitive to stellar population gradients,
metallicity variations, and dust attenuation than rest-frame optical measurements. Sizes correspond to circularised half-light radii,
$R_{1\mu{\rm m},50}$, and were obtained using \textsc{galfit} \citep{Peng2002} after combining archival HST data with new
JWST observations of the CANDELS fields \citep{Grogin2011,Koekemoer2011}. Only galaxies with reliable
\textsc{galfit} measurements and sizes exceeding the effective pixel scale are included.

The median relations and $16^{\rm th}$ to $84^{\rm th}$ percentile scatter are indicated by
connected circles and error bars, respectively. For \colibre, median relations are shown using different line styles, as in previous figures,
with shaded regions indicating the $16^{\rm th}-84^{\rm th}$ percentile scatter for L200m6. To facilitate a consistent
comparison, simulated galaxy sizes correspond to rest-frame $1\,\mu{\rm m}$ half-light radii, which we approximate using 
the 2D half-light radii measured in the GAMA $Y$-band (1.02\,$\mu{\rm m}$). These sizes were estimated from the corresponding surface brightness
profiles, calculated within 50 kpc apertures and for random orientations. Galaxies whose sizes fall below the observational resolution
limit of the \citet{Song2026} sample are excluded.

As in Fig.~\ref{fig:Size_Mass_vdWel}, we find good agreement between the simulated and observed size--mass relations at low and
intermediate redshifts, i.e. $z \lesssim 2$. At higher redshifts, systematic differences emerge. In particular, for 
$10^{9.5}\lesssim M_\star/{\rm M_\odot} \lesssim 10^{11}$, simulated galaxies are smaller than observed systems by
$\approx 0.1$--0.5~dex depending on their mass, while the size--mass relations at lower masses and for the highest-mass galaxies
remain in good agreement. While we
leave a comparison to virtual observations to future work, preliminary tests indicate that accounting for dust attenuation
increases the apparent sizes of intermediate-mass \colibre~galaxies.
At $z=3$, for example, the sizes of galaxies with $M_\star\gtrsim 10^9\,{\rm M_\odot}$ increase
by between $\approx 0.1$ -- 0.4 dex, with the largest increase occuring at $M_\star\approx {\rm a\,\,few}\times 10^{10}\,{\rm M}_\odot$,
which is comparable to the mass scale where we find the largest size differences between observed and simulated galaxies at this redshift
(see the lower panel, second from the right).

We have also compared the SMRs of star-forming and passive \colibre\ galaxies to those of \citet{Song2026},
although we do not show these results explicitly. Consistent with the trends seen in Fig.~\ref{fig:Size_Mass_vdWel}, we find that
at redshifts $z \gtrsim 1$ the separation between the SMRs of star-forming and passive galaxies is less pronounced
in \colibre\ than in the observations. In this regime, star-forming simulated galaxies are typically smaller than their observed
counterparts, while passive simulated galaxies are typically larger. This suggests that differences in the structure of
star-forming and quiescent systems at early times are somewhat muted in the simulations, or that the effects of dust should not
be ignored when modelling the dependence of the simulated SMR on star formation activity.\footnote{This result does not depend on
whether or not observational errors are added to the stellar masses.}

\begin{figure}
  \subfloat{\includegraphics[width=0.5\textwidth]{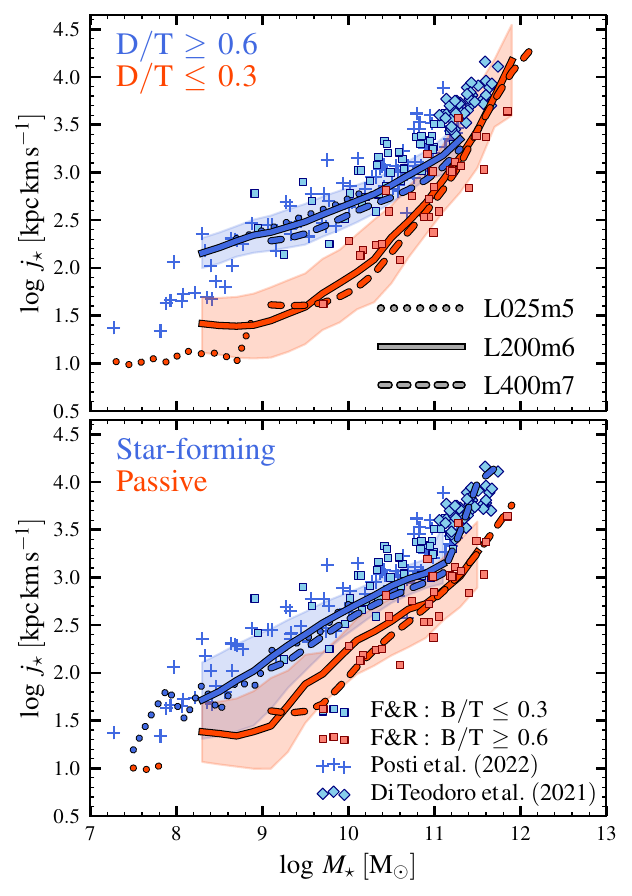}}
  \caption{Median stellar specific angular momentum ($j_\star$) as a function of stellar mass ($M_\star$) for simulated central
    galaxies at $z=0$. Early- and late-type systems are shown in blue and red, respectively. Dotted lines indicate L025m5, solid
    lines L200m6, and dashed lines L400m7. In the top panel, galaxies are classified by their kinematic disc-to-total ratio, $D/T$;
    in the bottom panel by star-formation activity (see text for selections). Observational measurements
    for disc galaxies are shown from \citet[][plus signs]{Posti2018} and \citet[][diamonds]{DiTeodoro2021}. Results from
    \citet{FR2018} are included for systems with $D/T\geq 0.6$ (blue squares) and $D/T\leq 0.3$ (red squares).
    For both classification schemes, \colibre~reproduces both the slope and normalisation of the observed $j_\star - M_\star$ 
    relations.} 
  \label{fig:jmstar}
\end{figure}

\subsection{The specific angular momentum--mass relation at $z=0$}
\label{sec:low-z-ang}

The results above show that \colibre's simulated galaxy population reproduces the observed stellar SMR, including its
systematic dependence on galaxy type, particularly for $z\lesssim 1$. Observational studies similarly reveal a strong morphological
segregation in the stellar specific angular momentum–mass plane, with star-forming discs and passive spheroids occupying distinct,
roughly parallel sequences \citep[e.g.][]{FR2018}. Motivated by this, we now extend our analysis to the stellar specific angular
momentum–mass relation, adopting the same morphological classifications that were used for galaxy sizes.

\subsubsection{The stellar angular momentum--mass relation}
\label{sec:JM-lowz}

In Fig.~\ref{fig:jmstar} we plot the median $z=0$ $j_\star - M_\star$ relations, using different line styles for the different resolution levels.
Galaxies are classified either by their kinematic disc-to-total ratio ($D/T$, top panel) or by specific star-formation rate (bottom panel). In the latter
case, star-forming galaxies lie within 1 dex of the main sequence (i.e. eq.~\ref{eqSFR}), while passive systems are at least 2 dex below it.
In both cases, the simulated population separates cleanly into two sequences that persist over the full mass range and exhibit
good convergence between resolutions. Late-type or actively star-forming systems define a higher-normalisation relation, while
early-type or passive galaxies populate a lower-$j_\star$ sequence.

We compare these results to observational measurements from \citet[][blue plus signs]{Posti2018}, \citet[][blue and red squares]{FR2018},
and \citet[][blue diamonds]{DiTeodoro2021}. The
SPARC-based measurements of \citet{Posti2018} and the massive spiral galaxies studied by \citet{DiTeodoro2021} trace the locus
of late-type systems in \colibre, while the morphologically-diverse sample of \citet{FR2018} brackets the early- and late-type sequences.

Overall, \colibre~reproduces both the slopes and relative normalisations of the observed $z=0$ $j_\star - M_\star$ relations.
The offset between late- and early-type systems, as well as the degree of intrinsic scatter---both of which
depend on stellar mass and sample selection---are consistent with observational
constraints, indicating that the simulated galaxies capture the strong morphology dependence of stellar angular momentum seen
in the local Universe. This mirrors the good agreement with the observed morphology-dependence of the SMRs shown
in Section~\ref{sec:SM-lowz}.

\subsubsection{The baryonic angular momentum--mass relation}
\label{sec:JMB-lowz}

\begin{figure}
  \subfloat{\includegraphics[width=0.48\textwidth]{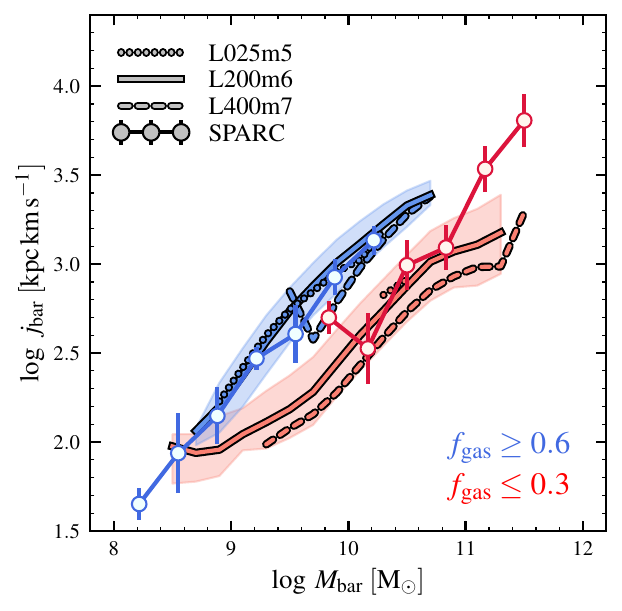}}
  \caption{The specific baryonic angular momentum--mass relation for $z=0$ late-type central galaxies (kinematic $D/T\geq0.6$) in \colibre,
    compared to SPARC galaxies from \citet{Zichen2025}. Dotted, solid, and dashed lines show the median relations for L025m5, L200m6, and L400m7,
    respectively, computed separately for systems with gas fractions $f_{\rm gas}\geq0.6$ (blue) and $f_{\rm gas}\leq0.3$ (red). Connected
    circles of matching colour show the corresponding median relations for SPARC galaxies selected using the same gas-fraction thresholds. The
    simulated relations are well converged and closely reproduce the observed trends, including their dependence on gas fraction.}
  \label{fig:jbar_Mstar}
\end{figure}

\begin{figure*}
  \subfloat{\includegraphics[width=0.9\textwidth]{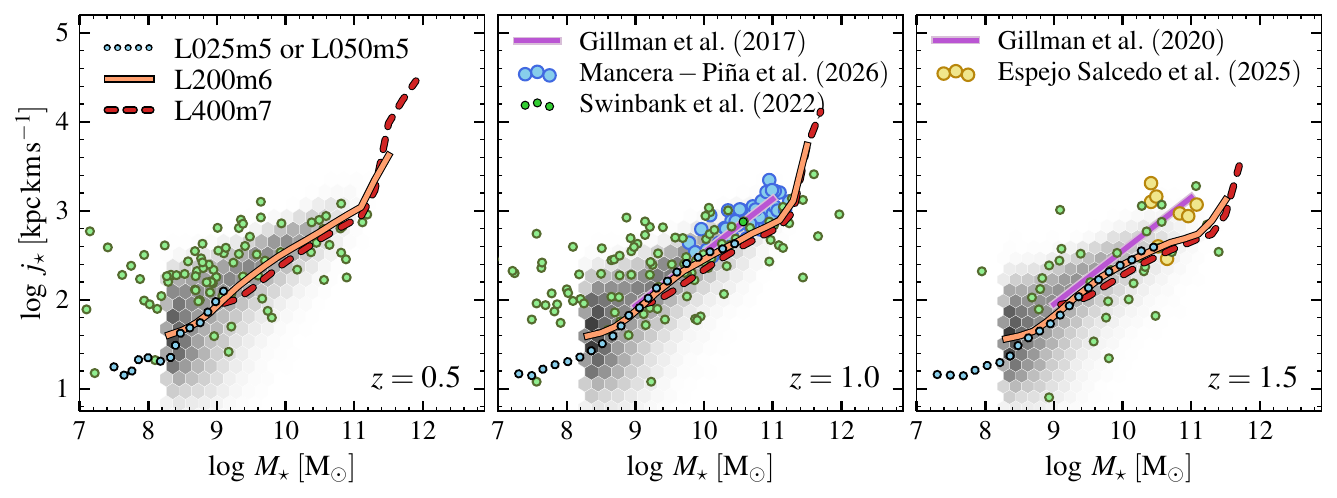}}
  \caption{The $j_\star$–$M_\star$ relations for central simulated galaxies at $z=0.5$, 1.0, and 1.5 (left to right, respectively) that
    lie within 1 dex of the redshift-dependent star-formation main sequence. The
    distributions of galaxies in L200m6 are shown as a 2D histogram; dotted, solid, and dashed lines indicate the
    median relations from L025m5 or L050m5, L200m6 and L400m7, respectively. In all panels, observational measurements from \citet{Swinbank2017}
    are shown as green points. In the middle panel, observations from \citet{Harrison2017} are shown as a purple line, and individual
    galaxies from \citet{PinaMancera2026} as blue circles. The right-hand panel shows the observed relation from \citet[][purple line]{Gillman2020}
    and individual galaxies from \citet[][yellow circles]{EspejoSalcedo2025} that cover the redshift range $1.4\leq z \leq 1.6$.
    The simulations demonstrate excellent convergence across resolution levels and align closely with observed relations.} 
  \label{fig:jmstar_highz}
\end{figure*}

Fig.~\ref{fig:jbar_Mstar} presents the baryonic angular momentum--mass relation for $z=0$ late-type (kinematic $D/T\geq0.6$) central galaxies
in \colibre. Median relations are computed separately for gas-rich ($f_{\rm gas}\geq0.6$) and gas-poor ($f_{\rm gas}\leq0.3$) systems, where
$f_{\rm gas}=1.33\,M_{\rm HI}/M_{\rm bar}$. The different \colibre~volumes are indicated using the same line styles as in Fig.~\ref{fig:jmstar}.

For comparison, we also show median relations for SPARC galaxies \citep{Zichen2025}, selected using the same gas-fraction thresholds and plotted
as connected circles with the same colour scheme as the simulations. At fixed gas fraction, galaxies follow approximately parallel relations in
the $\log\, j_{\rm bar}$--$\log \, M_{\rm bar}$ plane, with good convergence between the L025m5, L200m6, and L400m7 runs. All
simulated measurements remain broadly consistent with the normalization and scatter of the corresponding observational relations.

\vspace{0.5cm}
Overall, the angular-momentum content of $z\approx0$ central galaxies in \colibre~is in good agreement with observational
constraints, including its dependence on galaxy type, and independent of whether galaxies are classified by morphology or
star-formation activity, or whether angular momentum is defined for the stellar or total baryonic component. This mirrors
the level of agreement found for the various size–mass relations discussed in Sections~\ref{sec:SM-lowz} and \ref{sec:SM-highz}.

\subsection{Redshift evolution of the stellar specific angular momentum–mass relation}
\label{sec:JM-highz}

Fig.~\ref{fig:jmstar_highz} extends the comparison between simulated and observed $j_\star - M_\star$ relations to
redshifts $z\approx0.5$, 1.0, and 1.5. Grey hexagons show two-dimensional histograms of the $\log \, j_\star - \log \, M_\star$ distribution
for central star-forming galaxies (i.e. within 1 dex of the redshift-dependent star formation main sequence) in L200m6, while
solid orange lines indicate the median relation. Dotted blue and dashed red lines show the corresponding
medians for L050m5 and L400m7, respectively. At all redshifts, the relation exhibits an upturn at
$M_\star\gtrsim 10^{11}\,{\rm M_\odot}$, above which $j_\star$ rises more steeply with mass---though measurements in this
regime are particularly sensitive to aperture effects (see Appendix~\ref{app:aperture}).

Green points show observational measurements from \citet{Swinbank2017}, which are based on KMOS and MUSE observations of star-forming
main-sequence galaxies spanning $0.28\leq z\leq 1.65$. Galaxies with $0.28\leq z\leq0.75$ appear in the left, those with
$0.75<z\leq1.25$ in the middle, and those with $z>1.25$ in the right panel. At all redshifts, the simulated galaxies reproduce both the
normalisation and scatter of these observations.

The middle and right panels also include observational fits from \citet{Harrison2017} and \citet{Gillman2020}, derived from the
KROSS survey at $z\simeq0.9$ and the KGES survey at $z\simeq1.5$, respectively. In the middle panel, we additionally
show individual $z\simeq0.9$ galaxies from \citet{PinaMancera2026} as blue circles, and in the right panel we show
measurements for star-forming galaxies at $1.4\leq z \leq 1.6$ from \citet{EspejoSalcedo2025}. 

Overall, we find good agreement between simulated and observed $j_\star - M_\star$ relations across the full redshift range considered,
with most observational measurements lying within the intrinsic scatter of the simulated galaxy population. There is a hint that massive
simulated galaxies ($M_\star \gtrsim 10^{11}\,{\rm M_\odot}$) at $z=1.5$ have somewhat less angular momentum than observed systems, but the shape
of the relation in this regime is sensitive to scatter in $M_\star$, which tends to artificially flatten the relation (see Appendix~\ref{app:scatter}).

\begin{figure*}
  \subfloat{\includegraphics[width=0.9\textwidth]{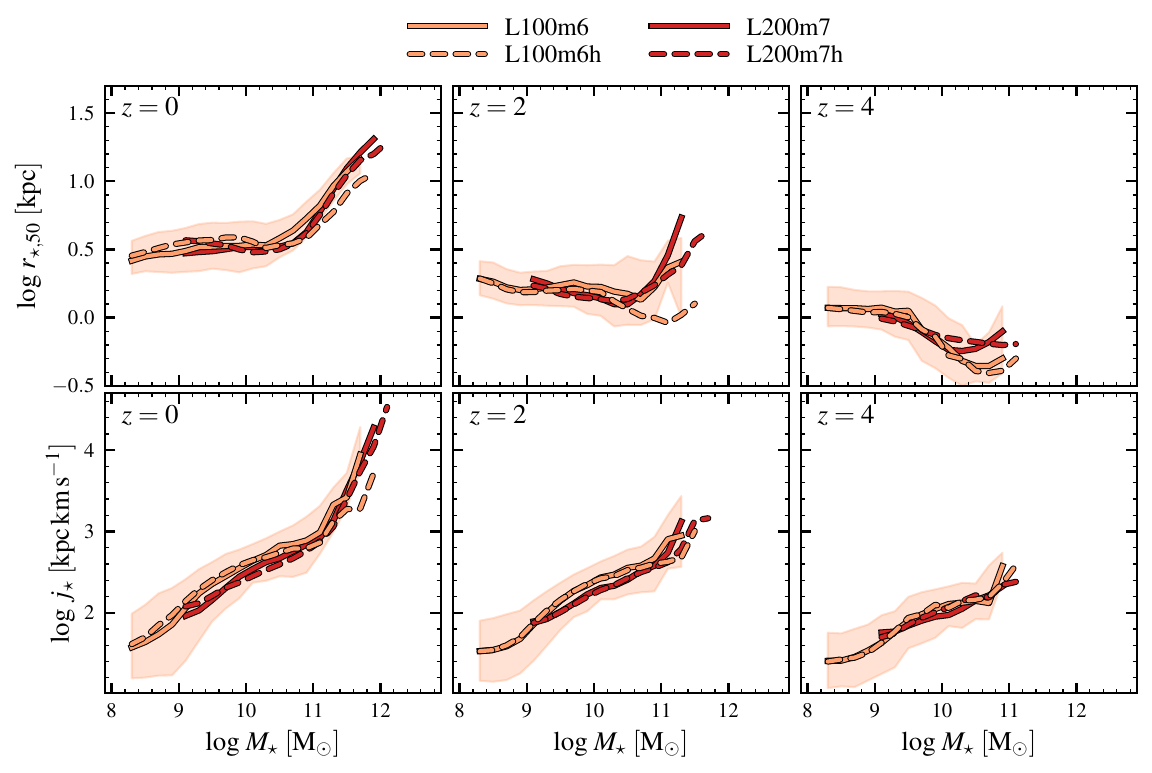}}
  \caption{Comparison of the $(100\,{\rm cMpc})^3$ m6 (yellow) and $(200\,{\rm cMpc})^3$ m7 (red) \colibre~volumes adopting the thermal- (solid lines)
    and hybrid-AGN (dashed lines) feedback schemes. Shaded regions show the $16^{\rm th}$ to $84^{\rm th}$ percentile scatter for galaxies in
    L100m6. Other runs exhibit comparable scatter, and are not shown for clarity. Columns correspond to $z=0$, $2$, and $4$. The top row shows
    the 3D size--mass relation for central galaxies with $M_\star\geq 100\times m_{\rm bar}$.
    Stellar masses and sizes are measured from bound stellar particles within $50\,{\rm kpc}$ of the galaxy centre.
    The bottom row shows the stellar specific angular momentum--mass relation for $M_\star\geq 100\times m_{\rm bar}$, where $j_\star$ is
    computed from all bound stellar material.
    The thermal- and  hybrid-AGN schemes produce similar galaxy properties at both resolutions.}
    \label{fig:ThermKin}
\end{figure*}

\subsection{Comparison of the thermal-AGN and hybrid-AGN feedback models}
\label{sec:hybrid}

All results presented in the preceding sections were obtained from runs that adopted the thermal-AGN feedback model. Here we compare those
predictions with an alternative hybrid scheme, in which thermal feedback is supplemented by kinetic jet-driven outflows. Both models were
calibrated at m7 resolution to reproduce the $z\approx0$ galaxy stellar mass function of \citet{Driver2022} and the size--mass relation
of \citet{Hardwick2022} over the stellar mass range $10^9 < M_\star/{\rm M_\odot} < 10^{11.3}$. Other galaxy properties, including angular
momentum, were not part of the calibration, nor was either model tuned to match $z>0$ observations. Differences between the two schemes
therefore provide insight into the sensitivity of galaxy structure to the details of AGN feedback.

In addition to their treatment of AGN feedback, the models differ slightly in their implementation of supernova feedback: the normalisation
of the gas-density-dependent heating temperature, $\Delta T_{\rm SN}(\rho_{\rm gas}) \propto \rho_{\rm gas}^{2/3}$, is reduced in the hybrid-AGN
model, resulting in lower heating temperatures at fixed $\rho_{\rm gas}$. This adjustment is intended to partially offset the stronger AGN
feedback in the hybrid scheme \citep[see][for details]{Husko2026}.

We present this comparison at two resolution levels: L100m6 and L100m6h, and L200m7 and L200m7h, the latter being the largest simulation
volume available for the hybrid-AGN scheme. This choice maximises the number of massive galaxies while ensuring comparable
statistics between the two volumes. Fig.~\ref{fig:ThermKin} summarises the comparison between the size--mass relations (upper panels)
and the specific angular momentum--mass relations (middle panels) at redshifts $z=0$, 2, and 4 (left to right).

At $z=0$, the the feedback schemes produce closely matching $r_{\star,50}$--$M_\star$ and $j_\star$--$M_\star$ relations, as expected
given the calibration. Small differences between the median relations of L200m7 and L200m7h lie well within the $16$th--$84$th percentile
scatter at all masses. The same is true for L100m6 and L100m6h, although there is some indication that the most massive galaxies at this resolution,
$M_\star\gtrsim10^{11.5}\,{\rm M_\odot}$, are $\approx0.1$--$0.2$ dex smaller and have correspondingly lower $j_\star$ in the hybrid-AGN model.

At $z=2$, the size--mass and angular momentum--mass relations remain very similar at low masses. In the m6 runs, however, galaxies with
$M_\star\gtrsim10^{10.5}\,{\rm M_\odot}$ are systematically larger in the thermal-AGN model (by $\approx 0.3$ dex), whereas in the m7 runs differences
appear only at $M_\star\gtrsim10^{11}\,{\rm M_\odot}$ (and typically remain within $\approx 0.2$ dex).
Above $M_\star\approx10^{11}\,{\rm M_\odot}$, galaxies in the thermal-AGN model also exhibit slightly higher specific angular momentum (by $\approx0.2$ dex),
although these offsets are confined to the most massive systems and are subject to small-number statistics.
These trends are consistent with more efficient removal of low-angular-momentum gas in the thermal-AGN model at these redshifts.

At $z=4$, the median size--mass and angular momentum--mass relations are again very similar between the two schemes, with each lying within the
scatter of the other. Differences between the models are therefore small across the full redshift range considered here.

Overall, the two feedback schemes produce very similar size--mass and angular momentum--mass relations, with only modest differences
emerging for the most massive galaxies at intermediate redshift. Taken together, these results indicate that the structural properties of galaxies
in \colibre~are reproduced consistently across resolution levels and AGN feedback implementations.

\section{Summary}
\label{Summary}

We have presented the stellar size--mass and specific angular momentum--mass relations for central galaxies in the
\textsc{Colibre} simulations \citep{Schaye2026,Chaikin2026a}, and compared them to a wide range of observational datasets. \textsc{Colibre}
improves upon previous cosmological simulations by offering: (i) a suite of volumes and resolutions that surpass 
previous simulation programs; (ii) a super-sampled dark
matter density field (four dark matter particles per baryonic particle), which reduces spurious energy transfer and improves the effective
resolution of baryonic structures; (iii) gas cooling below $10^4$~K, enabling star formation in dense and cold gas; and (iv) improved subgrid
models and a calibration strategy anchored to observed $z\approx0$ galaxy sizes, the stellar mass function, and black hole masses,
yielding improved convergence across resolutions.

The main results of our work are as follows:
\begin{enumerate}
\item The stellar size--mass relation exhibits a shallow slope below a characteristic mass 
  $M_\star\approx 10^{10.5}\,{\rm M_\odot}$.
  Above this mass, galaxy sizes increase strongly with stellar mass at $z\lesssim 2$, while at $z\approx 3$ they exhibit a 'U-shaped' trend,
  initially decreasing before increasing again at the highest masses. Below the characteristic mass, late-type galaxies are consistently
  larger (by $\approx 0.2$--$0.3$ dex) and have higher specific angular momentum than early types at fixed mass and redshift. Both populations
  follow approximately parallel $j_\star$--$M_\star$ relations, with offsets comparable to those observed between spiral and elliptical
  galaxies in the local Universe (Fig.~\ref{fig:Size_j_redshift}).

\item Galaxy sizes in \textsc{Colibre} are in excellent agreement with observations across a range of size definitions and samples.
  We find good agreement between 2D half-light radii derived from single-component S\'ersic fits to $r$-band surface brightness profiles and
  measurements from \citet{Casura2022}, as well as between 2D half-mass radii and the mass--size relations of
  \citet[][Fig.~\ref{fig:Size_Mass_Casura}]{Bellstedt2024}.
  This agreement extends to the $u$- and $z$-bands, and to the radii $R_{u,90}$ and $R_{z,90}$ enclosing
  90 per cent of the galaxy light in these bands (Fig.~\ref{fig:Size_GAMA_u_z}).

\item We also find very good agreement between the simulated $R_1$--$M_\star$ relations
  and the measurements of \citet{Trujillo2020} over 5 orders of magnitude in stellar mass,
  including when the relations are split by galaxy morphology (Fig.~\ref{fig:Size_Mass_Trujillo}).
  $R_1$ is the radius at which the stellar surface density falls below $\Sigma_1=1\,{\rm M_\odot\,pc^{-2}}$, and is typically
  larger than the projected half-mass radii use to calibrate \colibre's subgrid models by $\approx 0.2$ -- 1.0 dex, depending on
  stellar mass.

\item Similarly, the $z=0$ ``baryonic'' size--mass relations of late-type galaxies in \textsc{Colibre}, in which both stellar
  and gaseous components are included, agree very well with results from the SPARC survey \citep[][Fig.~\ref{fig:Rbar_Mstar}]{Zichen2025}.

\item At intermediate redshifts ($0.5\lesssim z\lesssim1.5$), the simulated 3D half-mass radii $r_{\star,50}$ are in good agreement with deprojected
  CANDELS size measurements from \citet[][Fig.~\ref{fig:Size_Mass_vdWel}]{vanderWel2024}. Comparisons of $1\,\mu{\rm m}$ half-light radii
  over the range $0\leq z\leq4$ to the observations of \citet{Song2026} also show good agreement at $z\lesssim1.5$. At higher redshifts
  ($z\gtrsim2$), galaxies with $M_\star\gtrsim 10^{9.5}\,{\rm M_\odot}$ are typically smaller than observed by $\approx 0.1$--$0.3$ dex,
  depending on mass (Fig.~\ref{fig:Size_Mass_Song}).
  While the size--mass relations of both star-forming and quenched galaxies agree well with observations at $z\lesssim1$, the separation
  between these populations becomes less pronounced than observed at higher redshift. Differences in the simulated
  and observed size--mass relations at high redshift are possibly due to the effects of dust attenuation, and will be addressed in future work.  

\item At $z=0$, late-type, star-forming discs have higher stellar specific angular momentum than quenched ellipticals, typically by
  $\approx 0.5$--$0.7$ dex depending on mass. Early- and late-type galaxies, classified using either their kinematic disc-to-total
  ratio or their offset from the star-forming main sequence, follow distinct $j_\star$--$M_\star$ relations with comparable
  scatter, in good agreement with observations of
  spirals and spheroids in the local Universe (Fig.~\ref{fig:jmstar}). The baryonic angular momentum--mass relation--defined using all
  baryons associated with a galaxy--also agrees well with observations of late-type galaxies, including its dependence on gas fraction
  (Fig.~\ref{fig:jbar_Mstar}).

\item The good agreement between the simulated and observed angular momentum--mass relations at $z=0$ extends at least to redshifts
  $z\approx 1.5$, with most available observational measurements lying within the intrinsic scatter of the simulated
  galaxy population  (Fig.~\ref{fig:jmstar_highz}). This mirrors the similarly strong agreement found for the size--mass
  relations over the same redshift range.
  
\item Our conclusions are robust to the adopted model for AGN feedback (Fig.~\ref{fig:ThermKin}) and to the exclusion of satellite galaxies
  from the analysis (Fig.~\ref{fig:A1}). However, the inferred sizes, stellar masses, and angular momenta (but not the kinematic morphology)
  do depend on the choice of spherical aperture within which these quantities are measured (Fig.~\ref{fig:A2} and \ref{fig:A3}).

\end{enumerate}

In summary, the \textsc{Colibre} simulations provide a self-consistent and observationally successful framework for studying the evolution
of galaxy sizes, angular momentum, and morphology across cosmic time. The close correspondence between simulated and observed scaling
relations—across multiple size definitions, redshifts, galaxy populations, and resolution levels—imples that \colibre~is capturing the
main physical processes that establish these relations; namely, feedback from stars and black holes. In addition, \textsc{Colibre} produces size--mass and angular-momentum--mass relations that
converge well across $\sim$2 orders of magnitude in stellar mass, including for both early- and late-type populations. This makes
\textsc{Colibre} a powerful tool for future studies aimed at disentangling the roles of angular momentum acquisition and retention,
feedback, mergers, and environment in shaping the galaxy population. We plan to address these issues in a follow-up paper.

\section*{Acknowledgements}
We thank Aaron Robotham, Jie Song, Zichen Hua, and Frederico Lelli for providing their observational
data and for helpful discussions. ADL acknowledges financial support from the Matariki Network of Universities.
DO is a recipient of an Australian Research Council Future Fellowship (FT190100083) funded by the Australian Government.
EC acknowledges support from STFC consolidated grant ST/X001075/1.
FH acknowledges funding from the Netherlands Organization for Scientific
Research (NWO) through research programme Athena 184.034.002. KAO acknowledges support by the Royal Society
through Dorothy Hodgkin Fellowship DHF/R1/231105. JT acknowledges support of a STFC Early Stage Research and
Development grant (ST/X004651/1). ABL acknowledges support by the Italian Ministry for Universities (MUR)
program ``Dipartimenti di Eccellenza 2023-2027'' within the Centro Bicocca di Cosmologia Quantitativa (BiCoQ),
and support by UNIMIB’s Fondo Di Ateneo Quota Competitiva (project 2024ATEQC-0050)
We also wish to acknowledge various public {\textsc{python}} packages that
have greatly benefited our work: {\textsc{scipy}} \citep{scipy}, {\textsc{numpy}} \citep{numpy},
{\textsc{matplotlib}} \citep{matplotlib} and {\textsc{ipython}} \citep{ipython}.
This work used the DiRAC Data Centric system at Durham University,
operated by the Institute for Computational Cosmology on behalf of the STFC DiRAC HPC
Facility (\href{www.dirac.ac.uk}{www.dirac.ac.uk}). This equipment was funded by a BIS National E-infrastructure
capital grant ST/K00042X/1, STFC capital grant ST/K00087X/1, DiRAC Operations grant
ST/K003267/1 and Durham University. DiRAC is part of the National E-Infrastructure.

\section*{Data Availability}
The \colibre~data underlying the figures and results presented in this paper are available from the
corresponding author upon reasonable request. A public version of the \textsc{Swift} simulation code
\citep{Schaller2024} is available at \url{http://www.swiftsim.com}. The \colibre-specific modules implemented
within \textsc{Swift} will be released at a later stage alongside the public simulation data.

\bibliographystyle{mnras}
\bibliography{paper} 

\appendix

\section{Convergence tests for galaxy sizes and angular momentum}
\label{app:convergence}

\begin{figure}
  \subfloat{\includegraphics[width=0.5\textwidth]{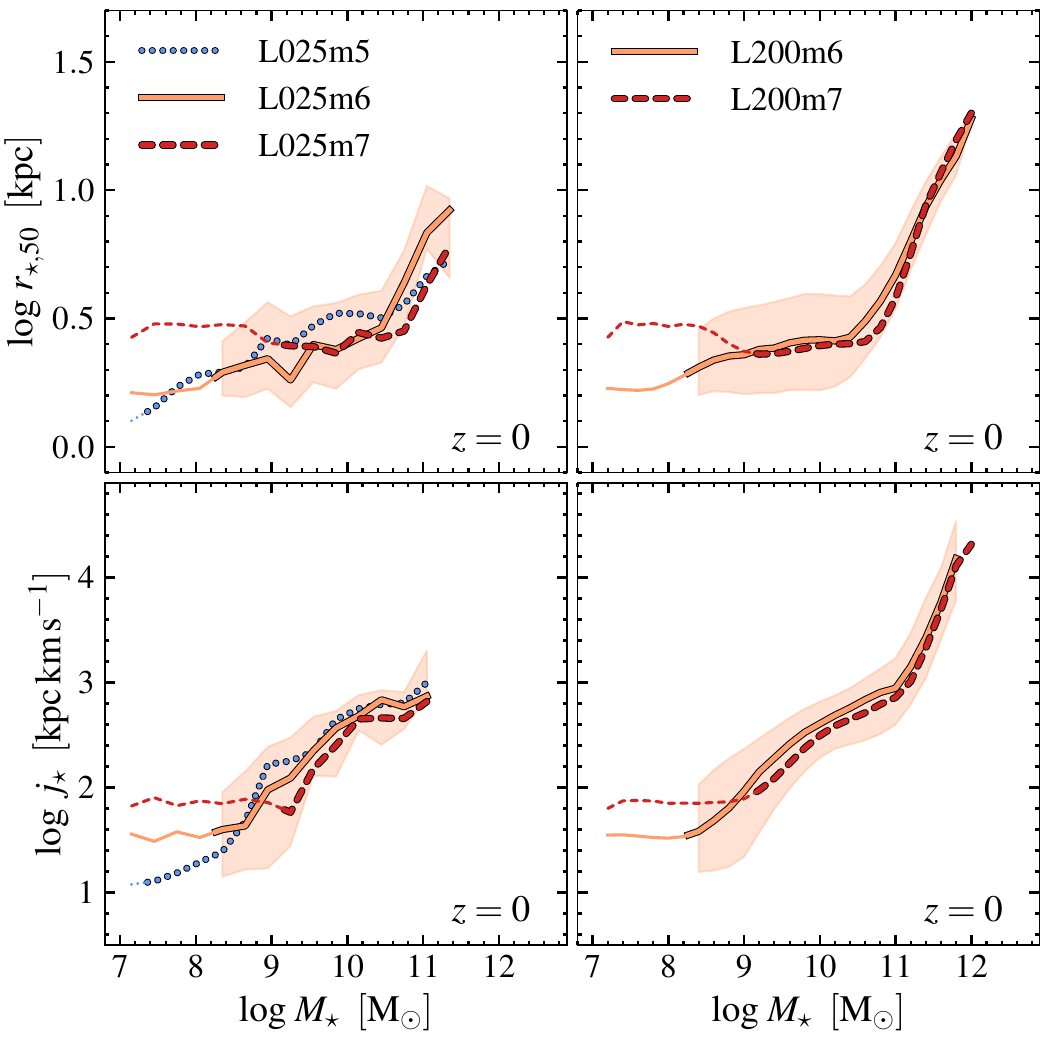}}
  \caption{Convergence of the three-dimensional stellar half-mass radii ($r_{\star,50}$;
    upper panels) and total stellar specific
    angular momentum ($j_\star$; lower panels) of central galaxies at $z=0$ in the thermal-AGN \colibre~simulations.
    Simulations of different numerical resolution are compared at fixed box size: $(25\,\mathrm{cMpc})^3$ volumes are shown in the
    left-hand panels, while $(200\,\mathrm{cMpc})^3$ volumes at the available resolutions are shown in the right-hand panels. The
    colour scheme and line styles follow those adopted throughout the paper. Thick lines span stellar mass bins from the nominal
    resolution limit, $M_\star=100\times m_{\rm bar}$ (where $m_{\rm bar}$ is the initial baryonic particle mass), up to the
    highest mass bin containing $\geq 10$ galaxies. Thin lines extend to lower stellar masses. Runs of different resolution
    exhibit good convergence for masses $M_\star\gtrsim 100\times m_{\rm bar}$.} 
  \label{fig:Aconv}
\end{figure}

Fig.~\ref{fig:Aconv} compares the $r_{\star,50}$--$M_\star$ (upper panels) and $j_\star$--$M_\star$ (lower
panels) relations for thermal-AGN models run with different numerical resolutions. Stellar half-mass radii, $r_{\star,50}$,
are measured directly from the particle data. To avoid mixing resolution effects with differences arising from simulation
volume, we compare runs of varying resolution at fixed box size. Results for the available $(25\,\mathrm{cMpc})^3$ volumes
are shown in the left-hand panels, while those for the $(200\,\mathrm{cMpc})^3$ volumes are shown in the right-hand panels.

For each set of curves, thick lines indicate stellar mass bins that both exceed $M_\star=100\times m_{\rm bar}$
(the nominal resolution limit adopted throughout this paper for quantities measured directly from particle
data) and contain $\geq 10$ galaxies. At lower masses, the relations are shown with thin lines.
We find that the sizes and angular momentum content of galaxies
resolved with $\gtrsim 100$ stellar particles converge well between simulations of different resolution. This resolution
threshold was therefore adopted in the preceding analysis. At lower masses, the curves deviate, with poorer resolution
corresponding to larger galaxy sizes and higher amounts of angular momentum. 

\section{Centrals versus satellites}
\label{app:centsat}

\begin{figure}
  \subfloat{\includegraphics[width=0.5\textwidth]{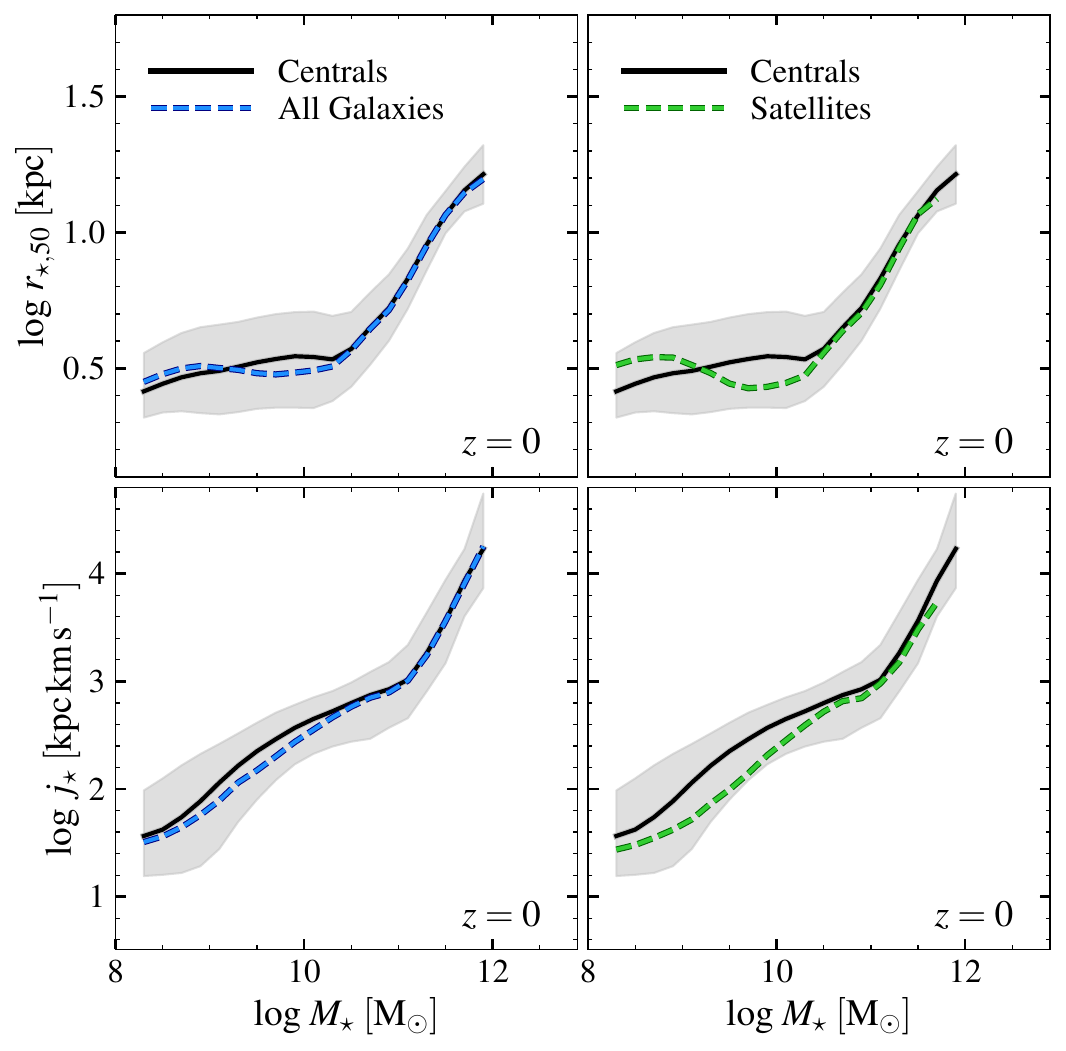}}
  \caption{The median $z=0$ $r_{\star,50}$--$M_\star$ (top panels) and $j_\star$--$M_\star$ relations
    (bottom panels) for different galaxy samples in the L200m6 (thermal-AGN) simulation. The left-hand panels compare
    central galaxies (solid black lines) with the full population of central and satellite galaxies (dashed blue lines).
    The right-hand panels compare central galaxies with satellites only (dashed green lines). Grey shaded regions
    indicate the 16$^{\rm th}$ to 84$^{\rm th}$ percentile scatter for central galaxies. Note that 
    the SMR and AMR for the full galaxy sample (centrals plus satellites) is very similar to the relations obtained for
    centrals alone.}
  \label{fig:A1}
\end{figure}

The results presented in Section~\ref{ref:results} focus exclusively on central galaxies in \colibre. This 
raises the possibility that the inferred relations between galaxy size, stellar angular momentum, and stellar mass
may be biased if satellite galaxies exhibit systematically different structural or kinematic properties due to, for
example, differences in their environmental evolution.

Fig.~\ref{fig:A1} compares the median $r_{\star,50}$--$M_\star$ (top panels) and
$j_\star$--$M_\star$ (bottom panels) relations for central galaxies at $z=0$ with those measured for the full
galaxy population (left-hand panels) and for satellites alone (right-hand panels) in L200m6. We find that the relations
derived for centrals alone are in close agreement with those obtained for the full population, indicating that restricting
our analysis to centrals does not significantly bias our results.

When centrals and satellites are considered separately, the correspondence remains good, although modest differences
are apparent. At $10^9 \lesssim M_\star/{\rm M_\odot} \lesssim 10^{10.5}$, satellite galaxies are, on average,
slightly more compact than centrals whereas lower-mass satellites are larger. For all $M_\star\lesssim 10^{11}\,{\rm M_\odot}$,
satellites have lower stellar specific angular momentum than centrals of the same mass. These offsets are small,
however, and do not qualitatively affect the trends discussed throughout this work.

The results plotted in Fig.~\ref{fig:A1} are limited to $z=0$, but we find similar, often better agreement between
the corresponding relations for centrals and satellites at higher redshifts. 

\section{The effect of aperture radius on galaxy size and angular momentum}
\label{app:aperture}

The stellar masses, sizes, andn $D/T$ ratios used in this paper were calculated using bound
stellar or gas particles within $50\,{\rm kpc}$ of the galaxy centre; stellar specific angular momentum was calculated using
all bound stellar material. Here we examine how these quantities depend on the choice of the three-dimensional
spherical aperture within which they are measured. Specifically, we compare the fraction of bound stellar mass
enclosed within apertures of radius $r_{\rm ap}=10$, 30, 50, and $100\,{\rm kpc}$ to the total bound stellar mass,
and assess the corresponding impact on galaxy size, angular momentum, and morphology. We focus on central
galaxies in the $(400\,{\rm cMpc})^3$ volume run with the thermal-AGN feedback model, which maximises the number of
massive galaxies for which aperture effects are expected to be most important.

\begin{figure}
  \subfloat{\includegraphics[width=0.5\textwidth]{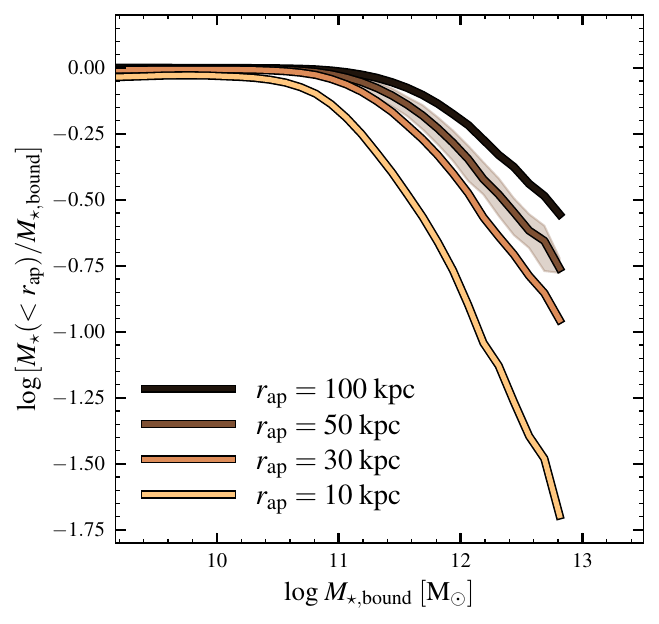}}
  \caption{Median fraction of bound stellar mass enclosed within different three-dimensional spherical apertures as a
    function of total bound stellar mass, $M_{\star,{\rm bound}}$. Results are shown for $z=0$ central galaxies in 
    L400m7 (thermal-AGN), which has the largest volume and therefore contains the highest number of massive galaxies
    for which aperture effects are most significant. For our fidcuial choice, $r_{\rm ap}=50\,{\rm kpc}$, we highlight the
    16$^{\rm th}$ to 84$^{\rm th}$ percentile scatter with a grey shaded region. While this fiducial measure captures nearly
    all bound stellar material for systems below $M_{\star,{\rm bound}} \lesssim 10^{11.5} \, {\rm M_\odot}$,
    an increasing fraction of bound mass resides beyond 50 kpc in more massive systems.}
  \label{fig:A2}
\end{figure}

Fig.~\ref{fig:A2} shows the fraction of bound stellar mass enclosed within each aperture as a function of total
bound stellar mass, $M_{\star,{\rm bound}}$. Below
$M_{\star,{\rm bound}}\approx 10^{10.5}\,{\rm M_\odot}$, all curves overlap, even for the smallest aperture considered
($r_{\rm ap}=10\,{\rm kpc}$), indicating that essentially all bound stellar material in galaxies below this mass
lies within $10\,{\rm kpc}$ of the galaxy centre. At higher masses the curves diverge, with smaller apertures
enclosing a progressively smaller fraction of the total bound stellar mass. For example, at
$M_{\star,{\rm bound}}=10^{12}\,{\rm M_\odot}$ only $\approx 15$ per cent of the bound stellar mass is contained within
$r_{\rm ap}=10\,{\rm kpc}$; this increases to $\approx 37$, 48, and 64 per cent for $r_{\rm ap}=30$, 50, and
$100\,{\rm kpc}$, respectively. Since galaxy sizes decrease with increasing redshift, the fraction of bound
stellar mass enclosed within fixed physical apertures is found to increase at earlier times (not shown explicitly here).

\begin{figure}
  \subfloat{\includegraphics[width=0.5\textwidth]{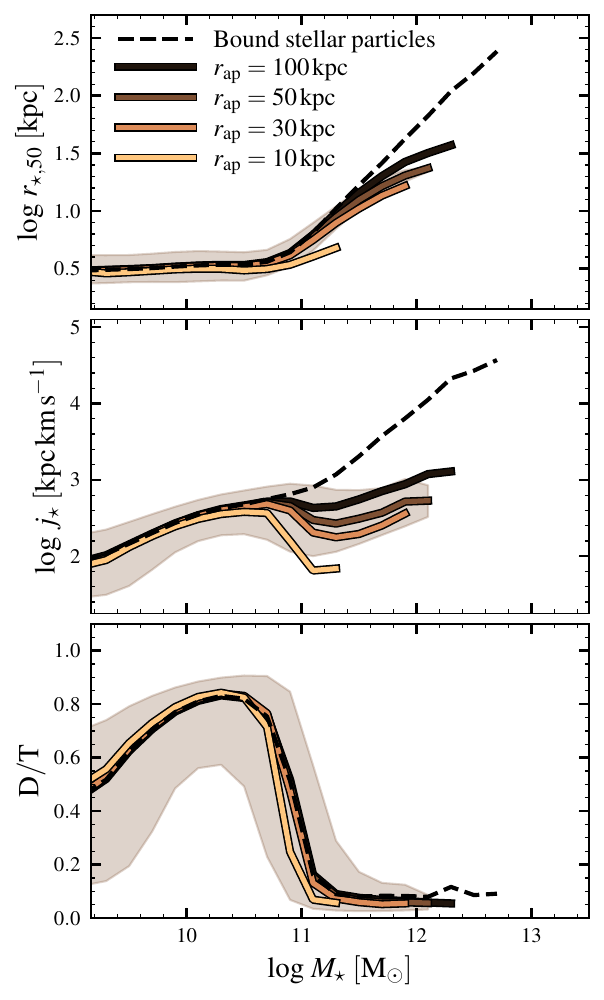}}
  \caption{The median $z=0$ $r_{\star,50}$--$M_\star$ (top), $j_\star$--$M_\star$ (middle), and
  $D/T$--$M_\star$ (bottom) relations for central galaxies computed using bound stellar particles
  within different three-dimensional apertures (coloured curves). Dashed black curves show the relations
  obtained using all bound stellar particles. Results are shown for L400m7 (thermal-AGN), which maximises the
  number of massive galaxies for which aperture effects are most pronounced. In all cases, quantities plotted on 
  the vertical and horizontal axes were computed using the same aperture. 
  For our fiducial choice ($r_{\rm ap}=50\,{\rm kpc}$), galaxy sizes for $M_\star \lesssim 10^{11.5}\,{\rm M_\odot}$ are
  consistent with results obtained using all bound stellar particles. Conversely, the specific angular momentum of all
  bound stellar particles exceeds that measured within any finite aperture considered for $M_\star \gtrsim 10^{11}\,{\rm M_\odot}$.
  The kinematic disc-to-total ratio remains largely insensitive to the choice of aperture across the entire mass range.}
  \label{fig:A3}
\end{figure}

Fig.~\ref{fig:A3} illustrates the impact of aperture choice on the $r_{\star,50}$--$M_\star$ (top),
$j_\star$--$M_\star$ (middle), and $D/T$--$M_\star$ (bottom) relations for central galaxies.
In all cases, quantities plotted on the vertical and horizontal axes were computed using the same aperture.
For comparison, relations obtained using all bound stellar particles are shown as dashed black curves.
Consistent with Fig.~\ref{fig:A2}, galaxy properties at
$M_\star\lesssim 10^{10.5}\,{\rm M_\odot}$ are insensitive to the choice of aperture.
At higher masses, larger apertures yield systematically larger galaxy sizes and higher specific angular momenta.

When all bound stellar particles are included, the size--mass relation is well described by a broken power law,
with a transition at $M_\star\approx 10^{10.7}\,{\rm M_\odot}$. Below this mass, median sizes are
approximately constant at $\approx 3\,{\rm kpc}$, while above it sizes increase steeply with stellar mass.
The angular momentum--mass relation obtained using all bound stellar material follows an approximate power law
for $M_\star\lesssim 10^{10}\,{\rm M_\odot}$, steepens for
$M_\star\gtrsim 10^{11}\,{\rm M_\odot}$, and exhibits a shallower slope at intermediate masses. This
curvature reflects the combined influence of the mass--morphology relation, the shape of the
stellar-to-halo mass relation, and the morphology-dependent normalisation of the
$j_\star$--$M_\star$ relation.

Throughout this paper, galaxy sizes and stellar masses are measured within a three-dimensional
$50\,{\rm kpc}$ aperture. For this choice, median sizes closely track those obtained using all bound stellar
particles for $M_\star\lesssim 10^{11}\,{\rm M_\odot}$. At higher masses, sizes measured within
$50\,{\rm kpc}$ are systematically smaller due to the exclusion of loosely bound stellar material associated
with extended stellar haloes. For the most massive galaxies, sizes based on a $50\,{\rm kpc}$ aperture can
differ from those measured using all bound stellar mass by up to $\approx 0.3$ dex. This agrees with the
results of \citet{Furlong2017} based on the \eagle\ simulation.

Specific angular momentum is considerably more sensitive to aperture choice than galaxy size, but again
only at high stellar masses ($M_\star\gtrsim 10^{10.7}\, {\rm M_\odot}$). This is because stellar
particles at large radii contribute disproportionately to the total angular momentum budget. Had the nominal
$50\,{\rm kpc}$ aperture been applied to the calculation of $j_\star$, the median specific angular momentum
would differ substantially from that obtained using all bound stellar material: by approximately 0.5 dex at
$M_\star\approx 10^{11.2}\,{\rm M_\odot}$, increasing to $\approx 1.2$ dex at
$M_\star\approx 10^{12}\,{\rm M_\odot}$. Aperture definitions should therefore be treated with care
when comparing the angular momentum content of simulated galaxies to that inferred from observations.

Finally, the kinematic $D/T$ ratios shown in the bottom panel of Fig.~\ref{fig:A2} are largely insensitive to the
choice of aperture. This implies that our selection of early- and late-type galaxies based on $D/T$ is robust
to aperture effects.

\section{The effect of redshift-dependent scatter in stellar masses}
\label{app:scatter}
\begin{figure*}
  \subfloat{\includegraphics[width=0.98\textwidth]{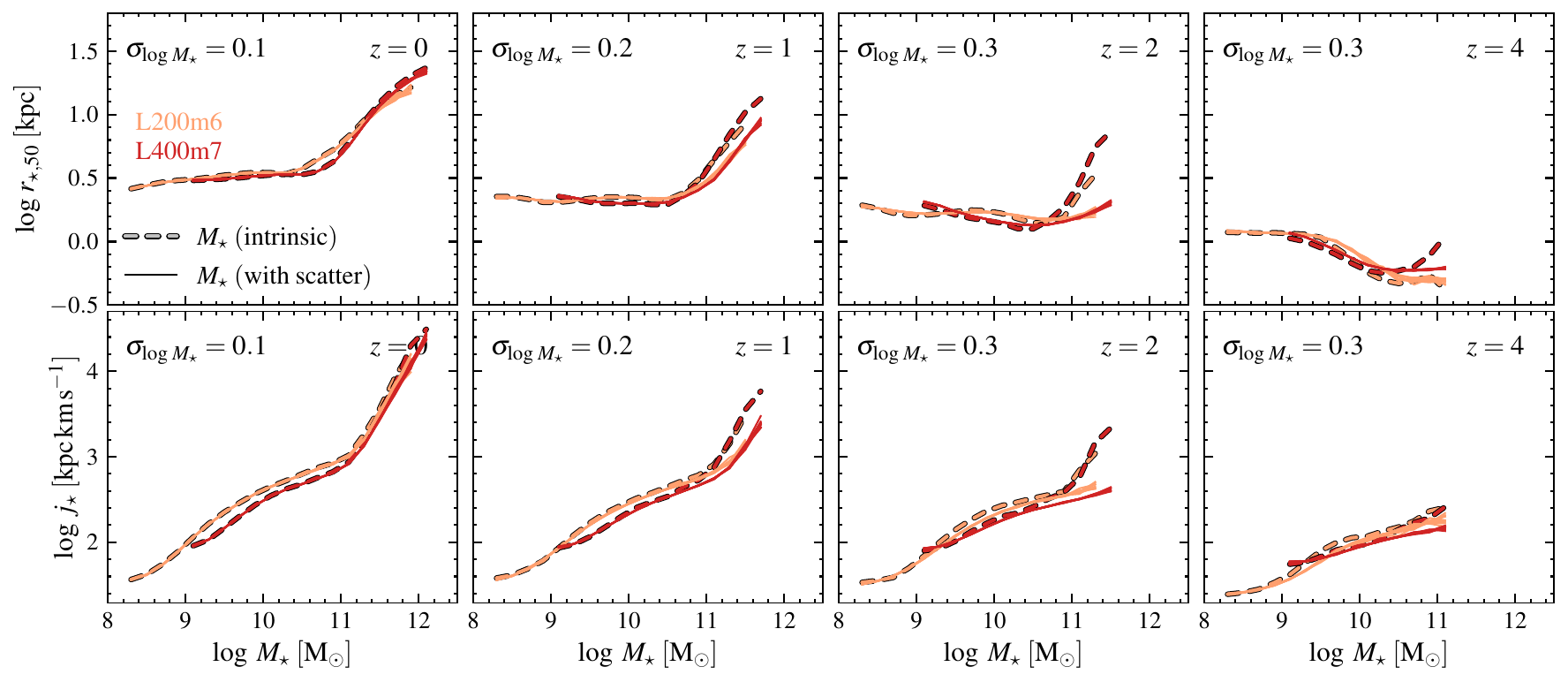}}
  \caption{The median $r_{\star,50}$--$M_\star$ (top) and $j_\star$--$M_\star$ (bottom) relations at $z=0$, 1, 2, and 4 (left to right, respectively). 
    Thick dashed lines show the intrinsic relations, where all quantities are measured directly from the simulated particle data. 
    Thin solid lines show the corresponding relations after adding a random lognormal scatter to the stellar masses and repeating this
    procedure 10 times. The standard deviation of $\log M_\star$ ($\sigma_{\log M_\star}$) is indicated in each panel. Orange and red lines show results
    for L200m6 and L400m7, respectively. When $\sigma_{\log M_\star}$ is small (leftmost panels), or when the relations are relatively
    flat, adding scatter to $M_\star$ has little effect. For steeper relations and larger $\sigma_{\log M_\star}$, the impact on the median
    relations can be significant (up to $\sim$0.5 dex at some masses and redshifts).}
  \label{fig:A4}
\end{figure*}

Fig.~\ref{fig:A4} shows the $r_{\star,50}$--$M_\star$ (top) and $j_\star$--$M_\star$ (bottom) relations at $z=0$, 1, 2, and 4 (left to right, respectively). 
Thick dashed lines represent the intrinsic relations measured directly from the simulated particle data. Thin lines show these relations after accounting
for observational errors by adding a redshift-dependent lognormal scatter, $\sigma_{\log M_\star}$ (see footnote~\ref{fn:obserr}). Ten realizations
of the scatter are shown in each case. For clarity, we only plot L200m6 (orange) and L400m7 (red); L025m5 and L050m5 yield consistent results.

The impact of scatter in $M_\star$ depends on both mass and redshift. Where the intrinsic relations are relatively flat (e.g. $r_{\star,50}$--$M_\star$ at
$M_\star \lesssim 10^{10.5}\,{\rm M_\odot}$ for $z \leq 2$, or $M_\star \lesssim 10^{9.5}\,{\rm M_\odot}$ at $z=4$), adding scatter has little
effect. In this regime, horizontal shifts in $M_\star$ do not significantly change the corresponding ordinate values, and the medians remain stable.

By contrast, when the intrinsic relations are steep and the scatter is large (e.g. both relations for $M_\star \gtrsim 10^{11}\,{\rm M_\odot}$ at $z \geq 1$), the median
relations becomes systematically shallower. This is due to the shape of the galaxy stellar mass function, which decreases towards higher masses. At fixed
$M_\star$, more galaxies scatter upward from lower masses—where sizes and angular momenta are typically smaller—than scatter downward from higher masses. 
This Eddington bias leads to shallower $r_{\star,50}$--$M_\star$ and $j_\star$--$M_\star$ relations. 
For intrinsically declining relations, the effect is reversed, and the slopes become steeper.

\section{Sersic indices from fits to stellar surface mass density profiles}
\label{app:nsersic}

\begin{figure*}
  \subfloat{\includegraphics[width=0.85\textwidth]{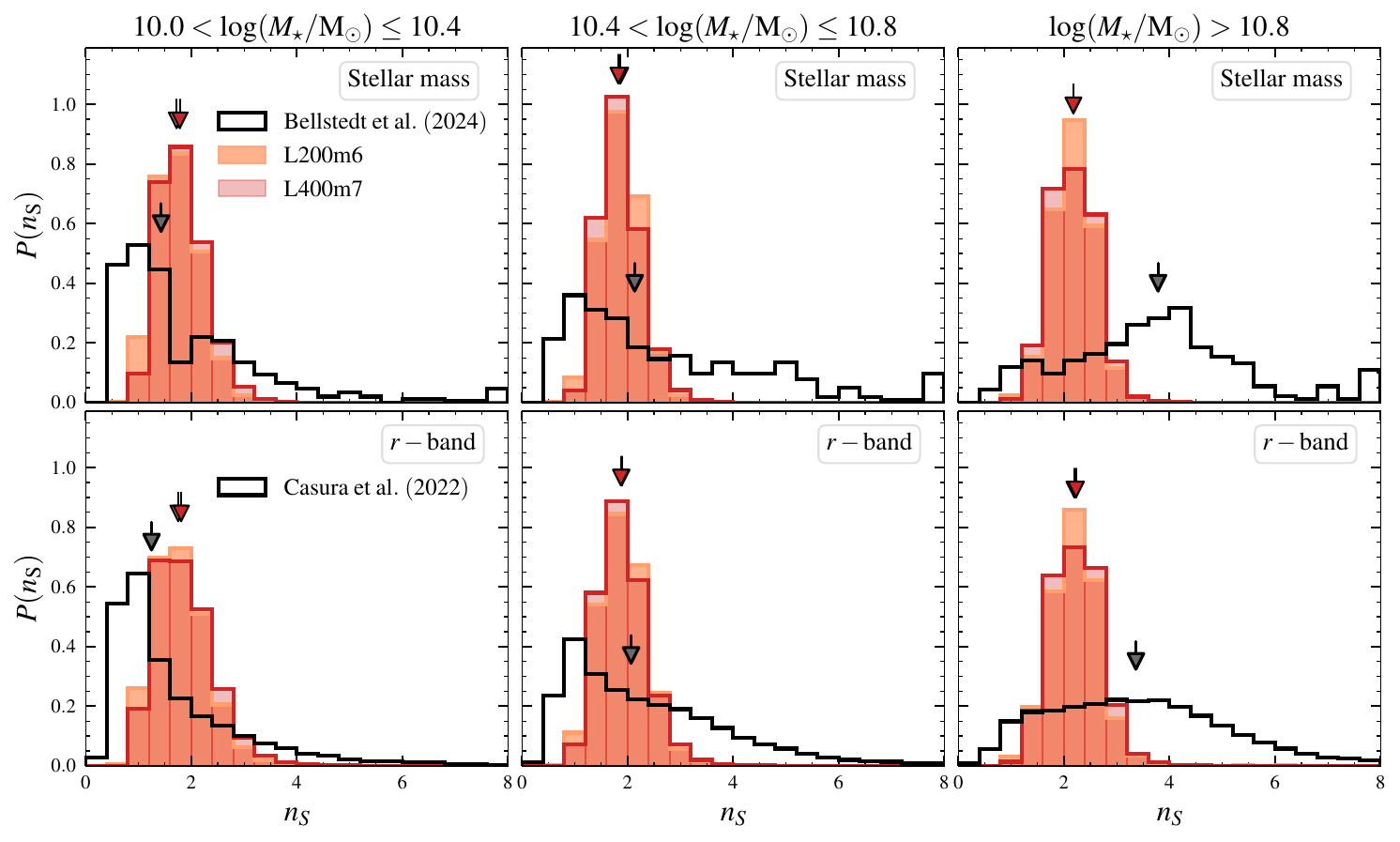}}
  \caption{Distributions of \sersic~indices for $z=0$ galaxies in bins of stellar mass. Top panels show results from fits to stellar mass surface density
      profiles, $\Sigma_\star$, while bottom panels show fits to $r$-band surface brightness profiles, $\Sigma_r$. Black histograms show all GAMA galaxies
      from \citet{Bellstedt2024} (top) and \citet{Casura2022} (bottom); orange and red histograms show central \colibre~galaxies from L200m6 and L400m7,
      respectively. The $\Sigma_\star$ and $\Sigma_r$ profiles of \colibre~galaxies exhibit relatively low \sersic~indices, with median values $n_{\rm S}\approx 2$
      and weak stellar mass dependence. In contrast, GAMA galaxies span a broader range of $n_{\rm S}$ and exhibit a stronger dependence on stellar mass.
      Morphologies inferred from \sersic~fits therefore suggest that \colibre~contains a lower fraction of bulge-dominated galaxies ($n_{\rm S}\approx 4$)
      than GAMA.}
  \label{fig:A5}
\end{figure*}

The \sersic~index characterizes the shape of surface brightness and stellar mass density profiles: late-type galaxies typically exhibit exponential profiles with
$n_{\rm S}\approx 1$, whereas early-type systems are better described by \citet{deVaucouleurs1948} profiles with $n_{\rm S}\approx 4$. Fig.~\ref{fig:A5} shows
distributions of \sersic~indices obtained from fits to stellar mass surface density profiles ($\Sigma_\star$; top panels) and unattenuated $r$-band surface brightness profiles
($\Sigma_r$; bottom panels) for GAMA galaxies and for $z=0$ central galaxies in L200m6 and L400m7.

The \colibre~galaxies exhibit a relatively narrow range of \sersic~indices, with weak stellar mass dependence. Median values increase from $n_{\rm S}\approx 1.7$ (1.8)
in the lowest stellar mass bin to $n_{\rm S}\approx 2.2$ (2.2) in the highest mass bin for the $\Sigma_\star$ ($\Sigma_r$) profiles, indicating that inferred
\sersic~indices are largely independent of whether they are measured from stellar mass or $r$-band light profiles.

In contrast, GAMA galaxies exhibit both a broader distribution of \sersic~indices and a stronger stellar mass dependence, with median values increasing from
$n_{\rm S}\approx 1.4$ (1.2) to $n_{\rm S}\approx 3.8$ (3.4) across the same stellar mass range for the $\Sigma_\star$ ($\Sigma_r$) profiles. Observationally
inferred values also depend on the imaging tracer, consistent with \citet{Kelvin2012}, who reported systematic differences between $n_{\rm S}$ values
inferred from optical and near-infrared imaging.

Similar discrepancies between simulated and observed \sersic~index distributions were reported for mock stellar mass and $r$-band imaging of \eagle~galaxies by
\citet{deGraaff2022}, who likewise found narrower distributions and a weaker stellar mass dependence than observed, but a stronger dependence on imaging technique
than found here for \colibre.

However, several methodological differences should be noted when comparing our results to observational measurements. Our $\Sigma_\star$ and $\Sigma_r$ profiles
are spherically averaged and do not include dust attenuation, instrumental effects, background noise, or point-spread-function convolution. In addition, we perform
one-dimensional fits to radial profiles, whereas observational studies typically fit two-dimensional \sersic~models, often allowing for radius-dependent ellipticity.
\sersic~indices can also be sensitive to the adopted fitting range. In our analysis, the inner fitting radius corresponds to the innermost bin enclosing at least 10
stellar particles, while the outer fitting radius corresponds to the outermost bin containing at least 10 stellar particles, or $50\,{\rm kpc}$ if larger. If the inner
regions are affected by numerical resolution---although Fig.~\ref{fig:Sigma} suggests this is not the case---or if the surface brightness at large radii falls below
the observational background level, then our fitting procedure may yield systematically different \sersic~indices from those inferred observationally. Nevertheless,
the similarity between our results and those of \citet{deGraaff2022} suggests that simulated galaxies may exhibit systematically different radial distributions of
stellar mass and light from those observed in real galaxies.

\bsp
\label{lastpage}
\end{document}